# Multivalued Logic Circuit Design for Binary Logic Interface

*A*

*Dissertation*

*Submitted*

*in partial fulfillment*

*for the award of the Degree of*

**Master of Technology**

**in Department of Computer Engineering**

**(With specialization in COMPUTER SCIENCE & ENGINEERING)**

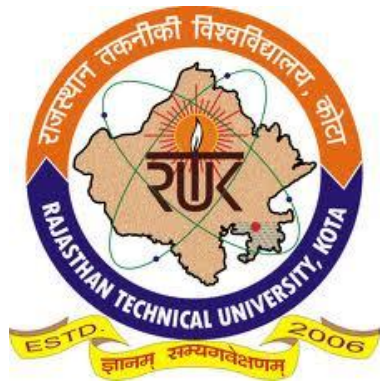

| | |
|---|---|
| **Supervisor** | **Submitted by** |
| Dr. S.C. Jain | Hitesh Gupta |
| Professor | Enrolment No.: |
| | 09E2UCCSM4XP606 |

**Department of Computer Engineering**

University College of Engineering

Rajasthan Technical University, Kota (Rajasthan)

**August – 2013**

i

# Candidate's Declaration

I hereby declare the work, which is being presented in the Dissertation, entitled **"Multivalued Logic Circuit Design for Binary Logic Interface"** in partial fulfillment for the award of Degree of "**Master of Technology**" in Department of Computer Engineering with specialization in Computer Science & Engineering, and submitted to the Department of Computer Engineering, University College of Engineering, Rajasthan Technical University is a record of my own investigation carried under the Guidance of **Dr. S.C. Jain**, Professor, Department of Computer Engineering, University College of Engineering, RTU Kota.

I have not submitted the matter presented in this Dissertation anywhere for the award of any other Degree.

**(Hitesh Gupta)**

M.Tech. (Computer Engineering)
Enrolment No.: 09E2UCCSM4XP606
University College of Engineering, RTU, KOTA

**Counter Signed by**

**Dr. S.C. Jain**

Professor
Department of Computer Science & Engineering
University College of Engineering
Rajasthan Technical University, Kota

# CERTIFICATE

       This is to certify that this Project entitled **"Multivalued Logic Circuit Design for Binary Logic Interface"** has been successfully carried out by **Hitesh Gupta** (Enrolment No.: 09E2UCCSM4XP606), under my supervision and guidance, in partial fulfilment of the requirement for the award of **Master of Technology** degree in Computer Science & Engineering from **University College of Engineering, Rajasthan Technical University, Kota** for year 2009-2011.

**Dr. S.C. Jain**

(Professor)
Department of Computer Engineering
University College of Engineering
Rajasthan Technical University, Kota

Place:   Kota

Date:

# ACKNOWEDGMENTS

I would like to take this opportunity to acknowledge and thank those who made this work possible and unforgettable to me.

First and foremost, thanks from bottom of my heart to the God, the omnipotent, for his blessing, showing me light even in a time of great darkness, and giving me strength. I would like to express my sincere gratitude to my Dissertation guide **Dr. S. C. Jain,** Professor in Department of Computer Science for his constant encouragement, able guidance and for giving me a new platform to build my career by giving me a chance to learn different fields of this technology. I am extremely grateful for what he has provided me.

I would also like to say thanks to my friend and research colleagues for their constant support and encouragement. I am extending my thanks to the M.Tech. students of university College of Engineering, Rajasthan Technical University, Kota for their invaluable support.

I express my deep sense of reverence to my parents and family members for their unconditional support, patience and encouragement.

Date:                                                                           **HITESH GUPTA**

# CONTENTS

# LIST OF TABLES

# LIST OF FIGURES

# LIST OF ALGORITHMS

# ABSTRACT

Binary logic and devices have been in used since inception with advancement and technology and millennium gate design era. The development in binary logic has become tedious and cumbersome. Multivalued logic enables significantly more information to be packed within a single digit. The design and development of logic circuits becomes very compact and easier. Attempts are being made to fabricate multivalued logic based devices.

Since present devices can be implemented only in binary system, it is necessary to evolve a system that can built the circuit in multivalued logic system and convert in binary logic system.

In multivalued logic system logic gates differ in different logic systems, a quaternary has become mature in terms of logic algebra and gates. Hence logic design based on above system can be done using standard procedure.

In this dissertation a logic circuit design entry based on multivalued logic system has been taken up that can provide the ease of circuit design in multivalued system and output as binary valued circuit. The named "MVL-DEV" offers editing, storage and conversion into binary facility.

<p style="text-align:right">**Chapter 1**</p>

# INTRODUCTION

Since inception digital devices have been designed using binary logic till date. Researchers have found the development in binary logic is cumbersome, complex and difficult to understand. Since multivalued logic enables more information to be packed in a single digit researchers have been working on multivalued logic for many years [1]-[3]. With development of novel electronic devices and optical devices, it is now possible to implements circuit for more complicated logic system [4]-[6].

Many of these devices are capable of dealing with more than two logic states but they are at experimental stage. Some multivalued logic systems such as ternary and quaternary logic schemes have been developed but successful implementation is yet to become available.

Quaternary logic has several advantages over binary logic. Since it require half the number of digits to store any information than its binary equivalent it is good for storage; the quaternary storage mechanism is less than twice as complex as the binary system. For the same reason, quaternary devices require simpler parallel circuit to process same amount of data than that needed in binary logic devices. Although several theoretical and physical variants of quaternary logic can already be found in [6]-[16].

The development of multivalued logic offers significant comfort over binary logic, but doesn't find target platform for implementation. We undertake this project to convert the multivalued logic to binary equivalent.

In this chapter section 1.1 describes about multivalued gates, section 1.2 describes about multivalued algebra, section 1.3 describes about multivalued conversion 1.4 describes about motivation section 1.5 describes about organization of dissertation.

## 1.1 Multivalued Gates

Logic gates constitute the foundation blocks for digital logic. Traditional binary gates are use in current digital devices. In multivalued logic gates have not been standardized but ternary and quaternary gates are two commonly used multivalued logic gates described in literature.

### 1.1.1 Binary Logic Gates

All the gates in this system can have two logic levels is input and output represented by 0, 1. Following are the some basic gates in this system.

An AND Binary gate has two or more inputs and produce one output as follows: output =1; if all the inputs are high, output=0; if one or more inputs are low [17].

Table 1.1: AND Gate Truth Table

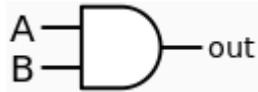

Fig. 1.1: AND Gate

| Input A | Input B | A.B |
|---------|---------|-----|
| 0 | 0 | 0 |
| 0 | 1 | 0 |
| 1 | 0 | 0 |
| 1 | 1 | 1 |

An OR Binary gate also has two or more inputs and produce one output as follows: output =1; if all the inputs are high, output=0; if one or more inputs are low [17].

Table 1.2: OR Gate Truth Table

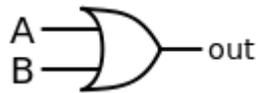

Fig. 1.2: OR Gate

| Input A | Input B | A+B |
|---------|---------|-----|
| 0 | 0 | 0 |
| 0 | 1 | 1 |
| 1 | 0 | 1 |
| 1 | 1 | 1 |

The inverter gate has one input and produces one output as follows: output =1; if input is low, output= 0; if input is high [17].

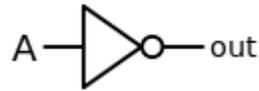

Fig. 1.3: NOT Gate

Table 1.3: NOT Gate Truth Table

| Input A | Output |
|---------|--------|
| 0 | 1 |
| 1 | 0 |

### 1.1.2 Ternary Logic Gates

Ternary logic system contain 3 input logic levels (0, 1, 2) different from quaternary and binary. In this multivalued logic system we have found some different gates. Some commonly used gates are discussed below and there symbols and truth tables are given below.

Table 1.4: LET 0 Gate Truth Table

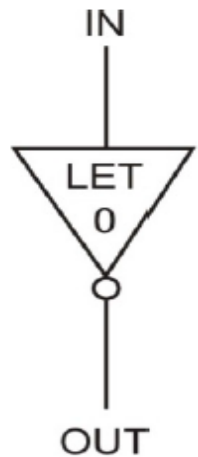

Fig. 1.4: LET 0 Gate

| IN | LET 0 |
|----|-------|
| 0 | 0 |
| 1 | 2 |
| 2 | 1 |

These gates are with one input and one output. There are few gates in this category like LET 1, LET 2, ROT 1, and ROT 2.

Now we will discuss gates with two inputs and one output. These gates based on concept when two quarrels and third win.

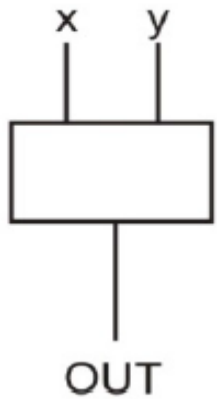

Fig. 1.5: Two quarrels and third wins Gate

Table 1.5: Truth Table of "Two quarrels and third wins gate"

| xy | OUT |
|----|-----|
| 00 | 0 |
| 01 | 2 |
| 02 | 1 |
| 10 | 2 |
| 11 | 1 |
| 12 | 0 |
| 20 | 1 |
| 21 | 0 |
| 22 | 2 |

Now the gate with three inputs and one output known as LIBRA gate.

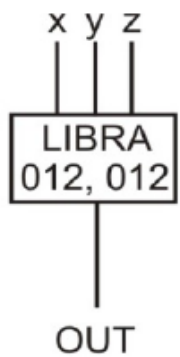

Fig. 1.6: LIBRA Gate 012,012

Table 1.6: Truth Table of LIBRA Gate 012,012

| xyz | OUT | xyz | OUT | xyz | OUT |
|-----|-----|-----|-----|-----|-----|
| 000 | 1 | 100 | 0 | 200 | 0 |
| 001 | 2 | 101 | 1 | 201 | 0 |
| 002 | 2 | 102 | 2 | 202 | 1 |
| 010 | 1 | 110 | 0 | 210 | 0 |
| 011 | 2 | 111 | 1 | 211 | 0 |
| 012 | 2 | 112 | 2 | 212 | 1 |
| 020 | 1 | 120 | 0 | 220 | 0 |
| 021 | 2 | 121 | 1 | 221 | 0 |
| 022 | 2 | 122 | 2 | 222 | 1 |

### 1.1.3 Quaternary Gate

As against binary gates have two logic levels {0,1}.quaternary gate has four logic levels{0, 1, 2, 3} in input and output.hence the following combinations could be {(0,0), (0,1), (0,2), (0,3), (1,1), (1,2), (1,3), (2,2), (2,3), (3,3) }available at input and corrsnding output according to algebrs define. Its truth table and gate symbol are given below:

AND Quaternary Gate:

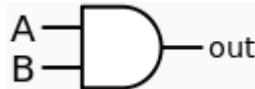

Fig. 1.7: Quaternary AND Gate

Table 1.7: Quaternary AND Gate Truth Table

| Operand | A | 0 | 0 | 0 | 0 | 1 | 1 | 1 | 2 | 2 | 3 |
|---------|---|---|---|---|---|---|---|---|---|---|---|
|         | B | 0 | 1 | 2 | 3 | 1 | 2 | 3 | 2 | 3 | 3 |
| Output | A.B | 0 | 0 | 0 | 0 | 1 | 0 | 1 | 2 | 2 | 3 |

OR Quaternary Gate:

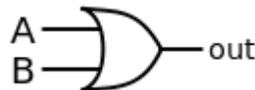

Fig. 1.8: Quaternary OR Gate

Table 1.8: Quaternary OR Gate Tuth Table

| Operand | A | 0 | 0 | 0 | 0 | 1 | 1 | 1 | 2 | 2 | 3 |
|---|---|---|---|---|---|---|---|---|---|---|---|
|  | B | 0 | 1 | 2 | 3 | 1 | 2 | 3 | 2 | 3 | 3 |
| Output | A+B | 0 | 1 | 2 | 3 | 1 | 3 | 3 | 2 | 3 | 3 |

As we can see here AND and OR gate symbol and their truth table. Like this we have more gate to be discussed furhter in this dissertation in chapter 3.

## 1.2 Multivalued Algebra

### 1.2.1 Boolean Algebra

Boolean algebra has similar rules to other algebras and these rules are used to manipulate the expression at hand. Some of these basic rules are [18].

One variable:

$$A . A = A$$
$$A + A = A$$

One variable and 0 or 1:

$$A . 0 = 0$$
$$A . 1 = A$$
$$A + 0 = A$$
$$A + 1 = 1$$

DeMorgan's Theorem:

$$(A + B)' = A' . B'$$
$$(A . B)' = A' + B'$$

Associative:

$$(A . B) . C = A . (B . C)$$
$$(A + B) + C = A + (B + C)$$

Commutative:

$$A \cdot B = B \cdot A$$
$$A + B = B + A$$

Distributive:

$$A \cdot (B + C) = A \cdot B + A \cdot C$$
$$A + (B \cdot C) = (A + B) \cdot (A + C)$$

### 1.2.2 Quaternary Algebra

In this section we will present fundamental properties of quaternary algebra. These properties are helpful to express and manipulate complicated functions algebraically to ensure efficient implementation. The packed-binary representation of quaternary digits and operators show that all fundamental operators except the bitswap obey the axioms and properties of Boolean operators. Most of these properties have their dual forms, where AND and OR operators are interchanged, at the same time the constants are inverted via NOT. The properties given below show that our proposed logic satisfies all the requirements to be algebra, as suggested by Huntington [19].

1) Closure

For every dyadic operator, $F(A, B) \in \{0, 1, 2, 3\}$, which is evident from definition. For every unary operator, $G(A) \in \{0, 1, 2, 3\}$.

2) Complement

There exists a unary operator NOT for which the following properties are true-

(1) $A + \bar{A} = <a_1 + \bar{a}_1, a_0 + \bar{a}_0> = <1, 1> = 3$

(2) $A \cdot \bar{A} = <a_1 \cdot \bar{a}_1, a_0 \cdot \bar{a}_0> = <0, 0> = 0$

Like these it also has some other properties of associativity, commutativity, distributivity, boundness. These properties would be discussed in detailed in further chapters 3.

## 1.3 Multivalued Conversion

As we have seen multivalued logic gates and their algebra are quite different ways for each binary, Quaternary, and Ternary. But the conversion between them is not straight forward.

The following example demonstrates the complexity of conversion. A simple equality gate in quaternary logic system is converted into three binary gates namely two XNOR and one AND gate. It target system does not allow XNOR. This XNOR require more number of gate and wiring connectivity.

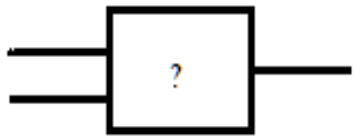

Fig. 1.9: Quaternary Equality Gate

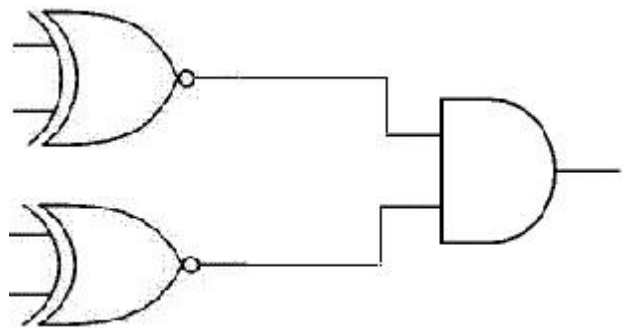

Fig. 1.10: Binary Conversion of Equality Gate

For large circuits, binary logic gate become very large and wiring adjustment becomes more complicated. A tool that can hide this complexity from users/ developers can be of immence use.

## 1.4 Objective and Motivation

As we feel that development in multivalued logic is comparatively, easier, faster and better understandable and likely to become new development environment but devices capable of implementing this logic have not yet become available. In order to take advantage of fast development a tool that can develop circuit in multivalued system and convert that in binary system for implementation purpose. Hence proposed tool will have the capabilities:

- ✓ Multivalued circuit schematic creation, editing , and saving.
- ✓ To integrate circuit components.
- ✓ Showing the every gate level property like truth table.
- ✓ Drawing of connectivity of gates.
- ✓ Binary conversion testing and binary converted circuit.

## 1.5 Organization of Dissertation

**Chapter 2:** Chapter named "LITERATURE SURVEY" will describes literature survey about multivalued valued crcuit which will help the readers to understand the basic and progress of different areas in field of multivalued logic.

**Chapter 3:** Chapter "MULTIVALUED LOGIC SYSTEM" is going to make planning of this dissertation work, it will genrate the basic understanding of multivalued logic gates, operators and their algebra. It is also explain the multivalued circuits and their truth table. Also quaternary logic system is discussed in this chapter.

**Chapter 4:** Chapter 4 with the name " DESIGN AND DEVLOPMENT OF INTERFACE TOOL" of this dissertation report is going to explain design decision and methodology with algorithm used to implement this work.

**Chapter 5:** This chapter name "IMPLEMENTATION AND RESULTS " is made for the purpose of graphical representaion and testing results.

**Chapter 6:** Finally this dissertation work is concluded with giving ideas of future extension of this dissertation work in this chapter with the name " CONCLUSION AND FUTURE SCOPE".

# Chapter 2

# LITERATURE SURVEY

As discussed earlier multivalued logic system is offering significant advantage like compact and easy development of circuits. A number of researchers have been working for evolution of multivalued logic system and complete algebra.

This chapter presents a survey of previous work in this area. The survey is organized in two major categories namely; multivalued logic system and multivalued logic algebra. Of course survey extension is given at the end of chapter.

## 2.1 Multivalued Logic System

Two logic systems namely ternary and quaternary have been developed and found in literature. A generic format named Galois field is also reported by researchers. We categorized our survey and these areas.

## 2.1.1 Ternary Logic System

Ternary logic system was reported in following literature: -

In 1990 X.W. Wu et al proposed CMOS compatible ternary logic based circuit and analyzed its implementation and difficulties. They describe that multiple power sources will be needed for multiple thresholds and a new theory of transmission function will be required [20].

In 2006 Xiaoqiang Shou, et al proposed a design methodology of latches with three stable operating points. Open-loop analysis is used to obtain insight into how a conventional binary latch structure can be modified to yield a ternary latch [21].

In 2010 Ion Profeanu proposed "A Ternary Arithmetic and Logic". This paper introduces ternary arithmetic and logic [22].

In 2010 Kanchan S. Gorde proposed The PTI and NTI have been designed using an inverter and pass-transistors at its output. The design of PTI and NTI is fully compatible with current CMOS technology [23].

In 2011 Afolayan A. Obiniyi, Ezugwu E. Absalom, Kwanashie Adako proposed "Arithmetic logic design with color coded ternary for ternary computing". This paper introduces a novel means of representing ternary states using color-codes, suggests a logic design model for a ternary half adder circuit and separate carry circuits. A ternary simulator was also developed to aid in the research and development of ternary systems [24].

In 2012 V.T.Gaikwad et al reviewed the development & advantages of ternary multi valued logic system and its CMOS implementation [25].

## 2.1.2 Quaternary Circuit

Quaternary logic system was reported in following literature: -

In 2000 "Takagi", et al suggested an extension of regular ternary logic function to function on discrete interval truth values. They presented this extension into quaternary valued logic and its mathematical properties [26].

In 2000 again Files et al have proposed multivalued decision diagrams (MDDs) similar to binary valued decision diagrams. For Boolean/multivalued input and output, completely/incompletely specified functions with application to logic synthesis, machine learning, data mining and knowledge discovery in databases [27].

In 2004 Khatri, et al proposed a PLA based Wire Removal techniques using SPFD method (sets of pairs of functions to be distinguished) for binary and multivalued circuits [28].

In 2005, Bundalo, et al have proposed a Design method for quaternary multiple valued combinational logic systems and circuits [29].

In 2010, Patel et al presented the design of a multiple-valued half adder and full adder circuits that are implemented in Multiple-Valued voltage-Mode Logic (MV-VML) [30].

In 2010, Cristiano Lazzari et al proposed a new FPGA structure based on a low-power quaternary voltage-mode device [31].

In 2010 Vasundara et al proposed a 4-bit counter using Multiple-valued D flip-flops and the proposed circuit is shown to exhibit less power dissipation [32].

In 2010 Vasundara et al presented arithmetic operations like addition, subtraction and multiplications in Modulo-4 arithmetic and in Galois field for the use in multi-valued logic design [33].

In 2012 Sheng Lin et al proposed a new novel design of a ternary memory cell using carbon nanotube field-effect transistors (CNTFETs) [34].

In 2012 Kumar et al studied the design, simulation of a CMOS quaternary logic generator having a single stage CMOS body [35].

Ifat Jahangir et al proposed but yet unpublished a new Quaternary algebra for implementing any quaternary circuit .In the paper a new minimal normal form for the quaternary algebra is proposed and the conversion of the quaternary logic into the binary is also stated [36].

## 2.1.3 Multivalued Logic

In 1993 D. Etiemble et al proposed a new algorithms for the sum of two (three and four) digits in the Binary Stored-Carry number system by using the smallest set of values for the positional sum. New adder is proposed using the same and the performance is evaluated against normal circuit [37].

In 1999 Elena Dubrova presented an overview of recent developments in multiple-valued logic circuit design, revealing both the opportunities they offer and the challenges they face [38].

In 1999 Brayton et al survey some of the methods used for manipulating, representing, and optimizing multi-valued logic with the view of both building a better understanding of the more specialized binary-valued logic, as well as motivating re-search towards a true multi-valued multi-level optimization package [39].

In 2000 Muthukrishnan et al, developed a multi-valued logic for quantum computing for use in multi-level quantum systems, and discuss the practical advantages of this approach for scaling up a quantum computer. The paper also generalizes the methods of binary quantum logic [40].

In 2004 A. Morgül et al proposed a new current-mode CMOS restoration circuit to restore or recover the nominal levels of the signal after a certain number of cascaded stages and the same is evaluated [41].

In 2011 Dhabhai et al performed the node reduction, area reduction and delay reduction by AIG rewriting [42].

## 2.2 Survey Extraction

In the above literature survey it is observed that multivalued circuits offer a great advantage over binary valued circuit in design size, effort and time. But implementation platform available is only binary.

Amongst the multivalued logic systems, quaternary logic system is well defined and offer significant advantage and design ease compared to others but in order to achieve compatibility with existing binary system (for implementation), an interfacing tool is necessary. In this project, we undertake development of such tool name "MVL-DEV". Subsequent chapters describe multivalued logic system and development of such tool.

**Chapter 3**

# MULTIVALUED LOGIC SYSTEM

In this chapter multivalued logic system is described. The system includes a set of logic gate, operators and algebra. The circuits can be designed using them. As discussed earlier quaternary logic system offers significant advantages in development. It is also a design entry method for our planned project. Hence quaternary logic system is discussed in subsequent sections.

First section explains various multivalued logic gates followed by multivalued logic algebra in middle section and then multivalued circuits are describes in last section.

## 3.1    Multivalued Logic Gates

In quaternary logic system, logic levels ranges from 0 to 3 as against 0 and 1 in binary logic. The logic system uses logic gates and their operation is known as operators. The gates and operators can be interchangeably used. The following quaternary gates/ operators are in use and have been considered in our project.

### 3.1.1    Quaternary Gates/ Operators

There are several operators in the quaternary algebra proposed in [36] are sufficient to describe any quaternary function. These operators are classified in two categories, namely fundamental operators and functional operators shown in Table 3.1.

Table 3.1: Classification of Quaternary Operators

| Quaternary Operators | |
|---|---|
| Fundamental Operators | Functional operators |
| AND, OR, NOT, BITSWAP | Inward Inverter, Outward Inverter, Equality, MIN, MAX, XOR |

**3.1.1.1 Fundamental Operators: -** Fundamental operators are those selected operators that are sufficient to completely define the quaternary algebra and can be used to derive other operators. They are defining below:

I.  AND GATE: - An AND gates have two inputs and produce one output in quaternary algebra. These gates are known as a fundamental operator. Their truth table and symbolic representation given below.

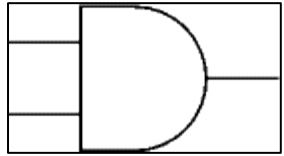

Fig. 3.1: Quaternary AND Gate

Table 3.2: Quaternary AND Gate Truth Table

| Operand | A | 0 | 0 | 0 | 0 | 1 | 1 | 1 | 2 | 2 | 3 |
|---|---|---|---|---|---|---|---|---|---|---|---|
|  | B | 0 | 1 | 2 | 3 | 1 | 2 | 3 | 2 | 3 | 3 |
| Output | A.B | 0 | 0 | 0 | 0 | 1 | 0 | 1 | 2 | 2 | 3 |

II. OR GATE: - An OR gates have two inputs and produce one output in quaternary algebra. These gates are also known as a fundamental operator. Their truth table and symbolic representation given below.

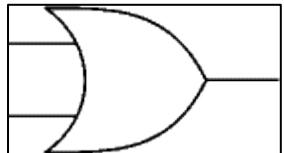

Fig. 3.2: Quaternary OR Gate

Table 3.3: Quaternary OR Gate Truth Table

| Operand | A | 0 | 0 | 0 | 0 | 1 | 1 | 1 | 2 | 2 | 3 |
|---|---|---|---|---|---|---|---|---|---|---|---|
|  | B | 0 | 1 | 2 | 3 | 1 | 2 | 3 | 2 | 3 | 3 |
| Output | A+B | 0 | 1 | 2 | 3 | 1 | 3 | 3 | 2 | 3 | 3 |

III. NOT GATE: - A NOT gates have one input and produce one output in quaternary algebra. These gates are also known as a fundamental operator. Truth table gives the output 0; in case of when input is 3.gives output 3; in case of input is 0. Gives output 2; in case of input is 1. And when output is 1; then input would be 2. Their truth table and symbolic representation given below.

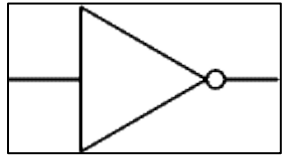

Fig. 3.3: Quaternary NOT Gate

Table 3.4: Quaternary NOT Gate Truth Table

| Operand | A | 0 | 0 | 0 | 0 | 1 | 1 | 1 | 2 | 2 | 3 |
|---|---|---|---|---|---|---|---|---|---|---|---|
| Output | $\overline{A}$ | 3 | 3 | 3 | 3 | 2 | 2 | 2 | 1 | 1 | 0 |

IV. BITSWAP GATE: - BITSWAP is only fundamental operator that does not have any binary equivalent and unique in algebra. It swaps the two bits of the binary equivalent of the quaternary operand. It leaves the symmetrical numbers unchanged but inverts the asymmetrical numbers. When the BITSWAP operator follows another operator, we get a derivative that might as well serve as an operator. Some example are BITSWAP AND (AND followed by BITSWAP), BITSWAP NOR (NOR followed by BITSWAP), BITSWAP XNOR (XNOR followed by BITSWAP) etc. In the BITSWAP NAND, NOR, NOT and XNOR the inverter is obviously NOT but not the inward or outward inverter. Figure 3.4 shows the BITSWAP operator, "~" is used in the symbol.

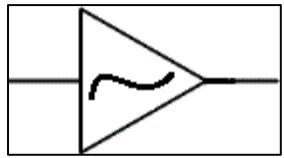

Fig. 3.4: Quaternary BITSWAP Gate

Table 3.5: Quaternary BITSWAP Gate Truth Table

| Operand | A | 0 | 0 | 0 | 0 | 1 | 1 | 1 | 2 | 2 | 3 |
|---|---|---|---|---|---|---|---|---|---|---|---|
| Output | ~ A | 0 | 0 | 0 | 0 | 2 | 2 | 2 | 1 | 1 | 3 |

**3.1.1.2 Functional operators**: - The functional operators are those operators that are described as the combination of two or more fundamental operators. It will be shown later that functional operators can also be used to express any arbitrary quaternary function [36]. Table 3.1 describes the operators as per this classification.

I. XOR GATE: - A XOR gates have two inputs and produce one output in quaternary algebra. These gates are known as a functional operator. Their truth table and symbolic representation given below:

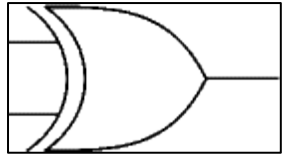

Fig. 3.5: Quaternary XOR Gate

Table 3.6: Quaternary NOT Gate Truth Table

| Operand | A | 0 | 0 | 0 | 0 | 1 | 1 | 1 | 2 | 2 | 3 |
|---|---|---|---|---|---|---|---|---|---|---|---|
| | B | 0 | 1 | 2 | 3 | 1 | 2 | 3 | 2 | 3 | 3 |
| Output | A⊕B | 0 | 1 | 2 | 3 | 0 | 3 | 2 | 0 | 1 | 0 |

II. INWARD INVERTER: - The inward inverter inverts the input just like the NOT; but after that, it changes the symmetrical values to nearest asymmetrical values. Therefore, when the output of the NOT would be 3 or 0, in the case of the inward inverter, the value will be 2 or 1, respectively. Inward inverters are called functional inverters. These inverters may as well be used with other operators to form compound operators; some examples are the inward AND, the inward OR and the inward XOR where the inward inverter follows the AND, OR and XOR operators respectively.

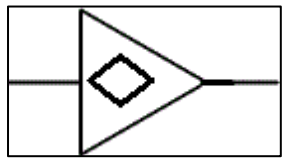

Fig. 3.6: Inward Inverter

Table 3.7: Inward Inverter Truth Table

| Operand | A  | 0 | 0 | 0 | 0 | 1 | 1 | 1 | 2 | 2 | 3 |
|---------|----|---|---|---|---|---|---|---|---|---|---|
| Output  | A' | 2 | 2 | 2 | 2 | 2 | 2 | 2 | 1 | 1 | 1 |

III. OUTWARD INVERTER: - The outward inverter inverts the input just like the NOT, but after that it changes the asymmetrical values to nearest symmetrical values. So when the output of the NOT would be 2 or 1, in the case of the outward inverter, the value will be 3 or 0, respectively. Outward inverters are called functional inverters. The operator is named as outward or full inverter because it converts the output of the NOT to the nearest absolute state. If we consider an outward inverter instead of the inward inverter in the above examples, then we get the outward AND, the outward OR and the outward XOR operators, respectively.

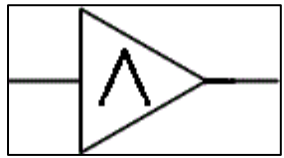

Fig. 3.7: Outward Inverter

Table 3.8: Outward Inverter Truth Table

| Operand | A  | 0 | 0 | 0 | 0 | 1 | 1 | 1 | 2 | 2 | 3 |
|---------|----|---|---|---|---|---|---|---|---|---|---|
| Output  | !A | 3 | 3 | 3 | 3 | 3 | 3 | 3 | 0 | 0 | 0 |

IV. EQUALITY: - The Equality operator is represented by a rectangular circuit symbol with a "?" inside it as shown in Fig. 3.8. We get the output in form of 0 and 3 only; when inputs are same than we get output 3, otherwise we get output 0.

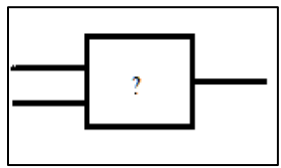

Fig. 3.8: Equality Gate

Table 3.9: Equality Gate Truth Table

| Operand | A | 0 | 0 | 0 | 0 | 1 | 1 | 1 | 2 | 2 | 3 |
|---|---|---|---|---|---|---|---|---|---|---|---|
|  | B | 0 | 1 | 2 | 3 | 1 | 2 | 3 | 2 | 3 | 3 |
| Output | E(A,B) | 3 | 0 | 0 | 0 | 3 | 0 | 0 | 3 | 0 | 3 |

V. **MAX GATE:** - MAX Gate is just like OR gate in representation and max is written inside the gate. Its symbol and truth table is given below.

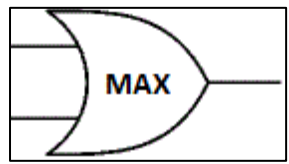

Fig. 3.9: MAX Gate

Table 3.10: MAX Gate Truth Table

| Operand | A | 0 | 0 | 0 | 0 | 1 | 1 | 1 | 2 | 2 | 3 |
|---|---|---|---|---|---|---|---|---|---|---|---|
|  | B | 0 | 1 | 2 | 3 | 1 | 2 | 3 | 2 | 3 | 3 |
| Output | A·B | 0 | 1 | 2 | 3 | 1 | 2 | 3 | 2 | 3 | 3 |

VI. **MIN GATE:** - MIN Gate is just like AND gate in representation and min is written inside the gate. Its symbol and truth table is given below.

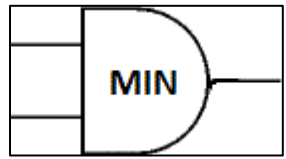

Fig. 3.10: MIN Gate

Table 3.11: MIN Gate Truth Table

| Operand | A | 0 | 0 | 0 | 0 | 1 | 1 | 1 | 2 | 2 | 3 |
|---|---|---|---|---|---|---|---|---|---|---|---|
|  | B | 0 | 1 | 2 | 3 | 1 | 2 | 3 | 2 | 3 | 3 |
| Output | A+B | 0 | 0 | 0 | 0 | 1 | 1 | 1 | 2 | 2 | 3 |

## 3.2   Multivalued Algebra

Quaternary algebra works on a set of operators and a set of values ranging from 0 to 3 for any input/output on logic gates. Quaternary digits {0, 1, 2, 3} can be imagined as 2-bit binary equivalents 00(absolute low), 01(intermediate low), 10(intermediate high), 11(absolute high). If the bits of the binary equivalent interchange their positions and still the quaternary state remains unchanged, then it is said to have binary symmetry; otherwise it is asymmetrical. It should be noted that both absolute states {0, 3} are symmetrical and both intermediate states {1, 2} are asymmetrical. A single quaternary digit is called a qudit when it is expressed as a number.

### 3.2.1   Quaternary Operators

A Quaternary digit can be expressed by two binary digits packed together using the following notion-

$$X = <x_1, x_0> = 2*x_1 + x_0 \qquad (1)$$

Where $x_1$ and $x_0$ are the constituent bits of the quaternary digit $X$ and the right side of (1) denotes the magnitude of $X$ in decimal system. A general fundamental operators (also known as dyadic operators) work like bitwise binary operators [36]-

$$F(X, Y) = F(x_1, x_0, y_1, y_0) = f(x_1, y_1), f(x_0, y_0) \qquad (2)$$

Where $F$ and $f$ stands for similar quaternary and binary operators respectively. The above notation of expressing quaternary digits (operators) in terms of binary digits (operators) is called packed-binary representation of quaternary digits (operators).

Using packed-binary representation, the NOT operator can be expressed quite easily in the following way -

$$\overline{X} = \overline{<x_1, x_0>} = <\overline{x_1}, \overline{x_0}> \qquad (3)$$

On the other hand BITSWAP can be expressed as-

$$\sim X = \sim <x_1, x_0> = <x_0, x_1> \qquad (4)$$

Binary bitswap, $\sim x \begin{cases} \overline{x} & ; x\ asymmetric \\ x & ; x\ symmetric \end{cases}$

The equality operator is defined as –

$$E(X\ Y) = E(Y\ X) = X^Y = Y^X = \begin{cases} 0 & ; \quad X \neq Y \\ 3 & ; \quad X = Y \end{cases} \quad (5)$$

Using packed-binary representation, the functional inverters can be expressed as

$$!X = !<x_1, x_0> = <\overline{x_1}, \overline{x_1}> \quad (6)$$

$$X' = <\overline{x_1}, \overline{x_0}>' = <\overline{x_1}, x_1> \quad (7)$$

From (6) and From (7), it can be written that,

$$!X = X' \oplus 1 \quad (8)$$

$$X' = !X \oplus 1 \quad (9)$$

### 3.2.2 Properties of Quaternary Algebra

The packed-binary representation of quaternary digits and operators show that all fundamental operators except the bitswap obey the axioms and properties of Boolean operators. Most of these properties have their dual forms, where AND and OR operators are interchanged, at the same time the constants are inverted via NOT. The properties given below show that logic satisfies all the requirements to be algebra, as suggested by Huntington [22].

a) *Closure*

For every dyadic operator, $F(A, B) \in \{0, 1, 2, 3\}$, which is evident from definition. For every unary operator, $G(A) \in \{0, 1, 2, 3\}$.

b) *Complement*

There exists a unary operator NOT for which the following properties are true-

(1) $A + \overline{A} = <a_1 + \overline{a}_1, a_0 + \overline{a}_0> = <1, 1> = 3$

(2) $A \cdot \overline{A} = <a_1 \cdot \overline{a}_1, a_0 \cdot \overline{a}_0> = <0, 0> = 0$

c) *Associativity*

(1) $A + (B + C) = <a_1 + (b_1 + c_1), a_0 + (b_0 + c_0)> = <(a_1 + b_1) + c_1, (a_0 + b_0) + c_0> = (A + B) + C$

(2) $A \cdot (B \cdot C) = <a_1 \cdot (b_1 \cdot c_1), a_0 \cdot (b_0 \cdot c_0)> = <(a_1 \cdot b_1) \cdot c_1, (a_0 \cdot b_0) \cdot c_0> = (A \cdot B) \cdot C$

d) *Commutativity*

 (1)  $A + B = <a_1 + b_1, a_0 + b_0> = <b_1 + a_1, b_0 + a_0> = B + A$

 (2)  $A . B = <a_1 . b_1, a_0 . b_0> = <b_1 . a_1, b_0 . a_0> = B . A$

e) *Distributivity*

 (1)  $A + (B.C) = <a_1 + (b_1.c_1), a_0 + (b_0.c_0)> = <(a_1 + b_1).(a_1 + c_1), (a_0 + b_0).(a_0 + c_0)> = (A + B).(A + C)$

 (2)  $A.(B + C) = <a_1 .(b_1+c_1), a_0.(b_0 + c_0)> = <(a_1. b_1) + (a_1. c_1), (a_0. b_0) + (a_0. c_0)> = (A . B) + (A . C)$

f) *Boundedness*

 (1)  $A + 0 = <a_1 + 0, a_0 + 0> = <a_1, a_0> = A$

    $A . 3 = <a_1 . 1, a_0 . 1> = <a_1, a_0> = A$

 (2)  $A + 3 = <a_1 + 1, a_0 + 1> = <1, 1> = 3$

    $A . 0 = <a_1. 0, a_0. 0> = <0, 0> = 0$

### 3.2.3 Properties of Quaternary Operators

*1) Bitswap operator distributes itself over AND and OR operators.*

$\sim (A + B) = \sim <a_1 + b_1, a_0 + b_0> = \sim <a_1, a_0> + \sim <b_1, b_0> = \sim A + \sim B$

$\sim A.B = \sim (<a_1, a_0>.<b_1, b_0>) = \sim <a_1, a_0>.\sim <b_1, b_0> = \sim A . \sim B$

*2) NOT obeys the De Morgan's law, when applied to the output of OR or AND gates.*

$\overline{A + B} = \overline{<a_1 + b_1, a_0 + b_0>} = <\overline{a_1}.\overline{b_1}, \overline{a_0}.\overline{b_0}> = \overline{A} . \overline{B}$

$\overline{A . B} = \overline{<a_1 . b_1, a_0 . b_0>} = <\overline{a_1}+\overline{b_1}, \overline{a_0}+\overline{b_0}> = \overline{A} + \overline{B}$

Using this property, AND, OR and NOT operators can be expressed in terms of NAND/NOR operator alone.

$\overline{A} = \overline{A.A} = \overline{A + A}$

$A.B = \overline{\overline{(A.B)}.\overline{(A.B)}} = \overline{(\overline{A + A}) + (\overline{B + B})}$

$A + B = \overline{\overline{(A.A)}.\overline{(B.B)}} = \overline{(\overline{A + B}) + (\overline{A + B})}$

The XOR and XNOR operators can be expressed in terms of NAND/NOR by using the above expressions of AND, OR and NOT.

*3) Like NOT, outward inverter also obeys the De Morgan's law, when applied to the output of OR or AND gates.*

$! (A+ B) = ! <a_1 + b_1, a_0 + b_0> = <\overline{(a_1 + b_1)}, \overline{(a_1 + b_1)}> = <\overline{a_1}, \overline{a_1}> . <\overline{b_1}, \overline{b_1}> =! A . ! B$

$! (A.B) = ! <a_1.b_1, a_0 .b_0> = <\overline{(a_1. b_1)}, \overline{(a_1. b_1)}> = <\overline{a_1}, \overline{a_1}> + <\overline{b_1}, \overline{b_1}> =! A+ ! B$

*4) There is no compact expression that can be used to express the distribution of inward inverter over AND or OR operators.*

$(A+ B)' = ( <a_1 + b_1, a_0 + b_0 >) = <\overline{(a_1 + b_1)}, a_1 + b_1 >'$

$(A.B)' = (<a_1.b_1, a_0 .b_0 >)' = <\overline{(a_1. b_1)}, a_1 . b_1 >$

None of the above can be expressed in a form similar to $<f ( a_1,b_1 ), f (a_0 ,b_0)>$.

The above arguments indicate that there is no algebraic expression to expand the operation of inward inverter following the AND or OR operation.

*5) The order of inward inverter and NOT can be reveresed.*

$(\overline{A})' = <\overline{a_1}, \overline{a_0}>' = <a_1 , \overline{a_1}> = <\overline{(a_1)}, a_1 > = (\overline{A'})$

*6) The order of outward inverter and NOT can be reversed.*

$!(\overline{A}) = ! <\overline{a_1}, \overline{a_0}> = <a_1, a_1> = <\overline{\overline{a_1}}, \overline{\overline{a_1}}> = (!\overline{A})$

*7) The order of bitswap and NOT can be altered.*

$\sim (\overline{A}) = \sim <\overline{a_1}, \overline{a_0}> = <\overline{a_0}, \overline{a_1}> = ! <a_0 , a_1> = (\overline{\sim A})$

*8) The order of bitswap and inward inverter can be altered under certain condition, not generally.*

$(\sim A)' = (<a_0 , a_1 >)' = <\overline{a_0}, a_0 >$

$\sim (A)' = \sim <\overline{a_1}, a_1> = <a_1 , \overline{a_1}>$

This implies, $(\sim A)' = \sim (A')$ if and only if $a_1 = \overline{a_0}$, i.e. $A$ is asymmetric.

9) *The order of bitswap and outward inverter can be altered under certain condition, not generally.*

$$\sim(!\ A) = \sim<\overline{a_1}, \overline{a_1}> \ = <\overline{a_1}, \overline{a_1}>$$
$$!\ (\sim A) = !<\overline{a_0}, \overline{a_1}> = <\overline{a_0}, \overline{a_0}>$$

This implies, $\sim (!\ A) = !\ (\sim A)$ if and only if $a_1 = a_0$, i.e. $A$ is symmetric.

10) *The order of inward and outward inverters can never be reversed under any condition.*

$$(!A)' = (<\overline{a_1}, \overline{a_1}>)\ ' = <a_1, \overline{a_1}>$$
$$!(A') = !<\overline{a_1}, a_1> = <a_1, a_1>$$

This implies, $(!\ A)' \neq !\ (A')$ under any circumstances.

### 3.2.4 Theorems of Quaternary Algebra

There are several theorems in quaternary algebra that are derived from the fundamental postulates of the algebra and properties of the operators. Here we present a list of theorems that are useful in algebraic operations –

*1) The Law of Idempotency:*

$$X + X = X, \qquad X \cdot X = X$$

*2) The Law of Absorption:*

$$X + (X \cdot Y) = X, \qquad X \cdot (X + Y) = X$$

*3) The Law of Identity:*

$$X + Y = X, \quad X \cdot Y = X; \qquad \text{for } X = Y$$

*4) The Law of Complements with NOT:*

$$X + Y = 3, \quad X \cdot Y = 0; \qquad \text{for } X = \overline{Y}$$

*5) The Law of Involution with NOT and bitswap:*

$$X = \overline{\overline{X}}, \qquad X = \sim (\sim X)$$

*6) The Law of Elimination with NOT:*

$$X + \overline{X} \cdot Y = X + Y, \qquad X \cdot (\overline{X} + Y) = X \cdot Y$$

*7) The Law of Concensus with NOT:*

$$X \cdot Y + \overline{X} \cdot Z + Y \cdot Z = X \cdot Y + \overline{X} \cdot Z$$
$$(X + Y) \cdot (\overline{X} + Z) \cdot (Y + Z) = (X + Y) \cdot \overline{X} + Z)$$

*8) The Law of Interchange with NOT:*

$$(X.Y) + (\overline{X}.Z) = (X + Y).(\overline{X} + Z)$$

$$(X + Y).(\overline{X} + Z) = (X.Z) + (\overline{X}.Y)$$

### 3.3 Multivalued Circuits

In this sub-section, some quaternary valued circuits have been described. There are few circuit added and their quaternary truth table given respectively. Their equivalent binary conversion is explained further.

Table 3.11: Multivalued Circuit

| S.N | Circuit name | Schematic | Truth Table |
|---|---|---|---|
| 1 | Quaternary 1 to 4 decoder | (schematic: S input to four ? blocks labeled 0,1,2,3 producing L[0], L[1], L[2], L[3]) | S \| Decoder (L) [0] [1] [2] [3]<br>0 \| 3 0 0 0<br>1 \| 0 3 0 0<br>2 \| 0 0 3 0<br>3 \| 0 0 0 3 |
| 2 | Quaternary 1 to 4 de-multiplexer | (schematic: S into DECODER, D input, AND gates producing L[0], L[1], L[2], L[3]) | S \| Demultiplexer (L) [0] [1] [2] [3]<br>0 \| D 0 0 0<br>1 \| 0 D 0 0<br>2 \| 0 0 D 0<br>3 \| 0 0 0 D |

| 3 | A quaternary 4X1 multiplexer using a 1X4 decoder | 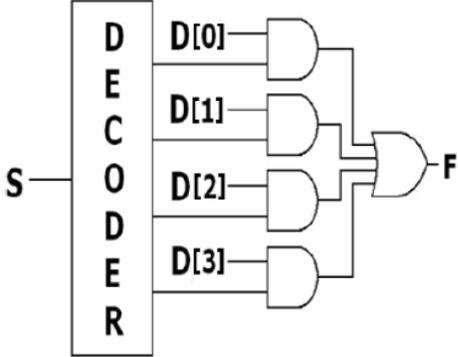 | S | Multiplexer (F) |
|---|---|---|---|---|
| | | | 0 | $D_0$ |
| | | | 1 | $D_1$ |
| | | | 2 | $D_2$ |
| | | | 3 | $D_3$ |

| 4 | Arbitrary Circuit #1 | 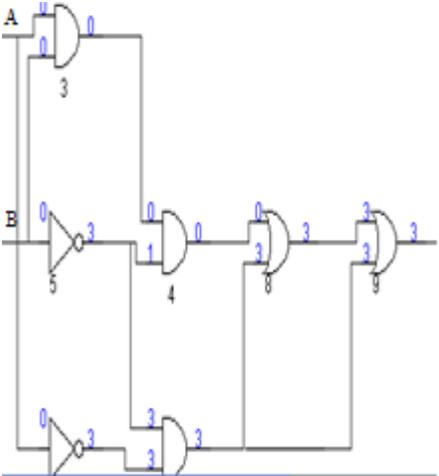 | A | B | Y |
|---|---|---|---|---|---|
| | | | 0 | 0 | 3 |
| | | | 0 | 1 | 2 |
| | | | 0 | 2 | 1 |
| | | | 0 | 3 | 0 |
| | | | 1 | 0 | 2 |
| | | | 1 | 1 | 3 |
| | | | 1 | 2 | 0 |
| | | | 1 | 3 | 0 |
| | | | 2 | 0 | 1 |
| | | | 2 | 1 | 0 |
| | | | 2 | 2 | 3 |
| | | | 2 | 3 | 0 |
| | | | 3 | 0 | 0 |
| | | | 3 | 1 | 1 |
| | | | 3 | 2 | 2 |
| | | | 3 | 3 | 0 |

| 5 | Arbitrary Circuit #2 | 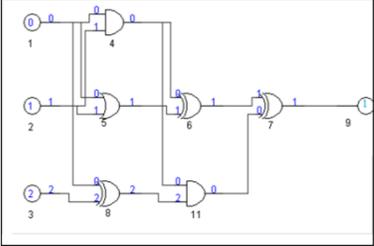 | Truth Table is given further in chapter 5 |
|---|---|---|---|

The tool has been planned to take any quaternary valued circuit in input. Some of the circuits have been used in testing the tool and described detailed in chapter 5.

<div style="text-align: right;">Chapter 4</div>

# DESIGN AND DEVLOPMENT OF INTERFACE TOOL

This chapter describes design and development of logic design entry tool named "MVL DEV". Objective of this tool is to capture multivalued logic based design and work as a plugin for binary value system.

The tool will provide multivalued circuit development environment. This chapter is going to explain Design flow, GUI Design and conversion algorithm. This tool will be discussed in subsequent sections.

## 4.1    Design Flow

In order to obtain binary logic based circuit, there are two approaches considered for conversion. Design flow of both is described in fig 4.1 (a) and 4.1 (b).

A former approach is based on schematic design entry, followed by replacement based conversion. Whereas later approach converts at truth table level and synthesize binary valued circuit from truth table.

Replacement based conversion of the circuit starts replacing circuit components from beginning and continues replacement till end. The algorithm then optimizes the circuit wherever required and setup connections within binary system.

Later approach converts the circuit in quaternary truth table and then converts into binary truth table. Using logic synthesis process, binary circuits can be drawn.

Former approach can be helpful in incremental design but later cannot be used till full design is complete. Hence we have considered former approach in our development.

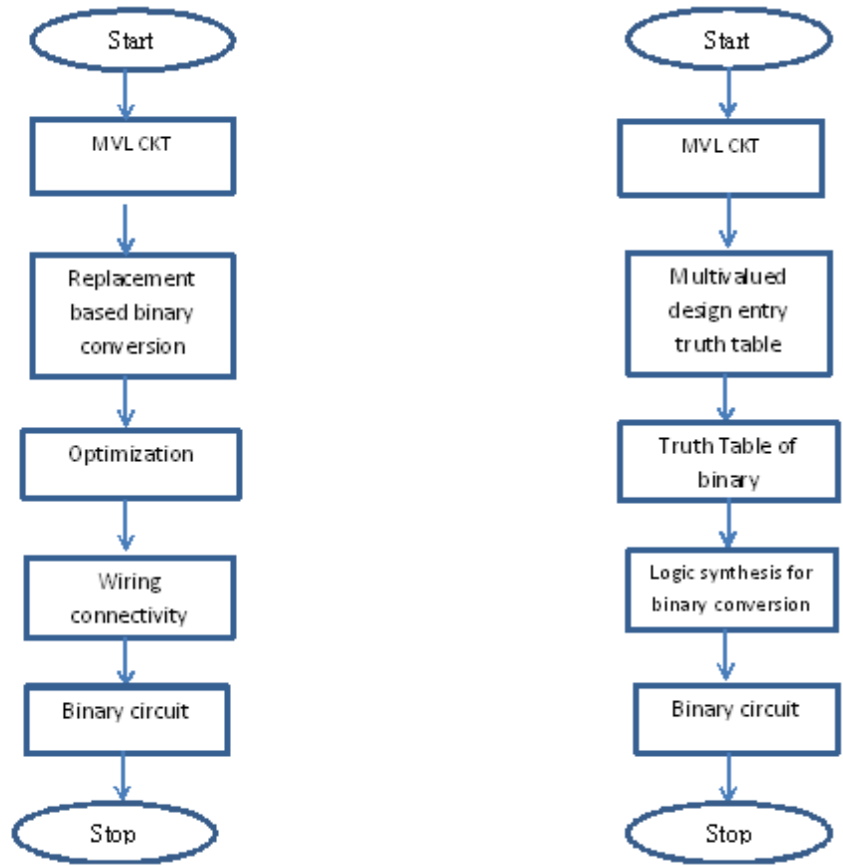

Fig. 4.1: (a)　　Replacement based approach　　　　　　(b) Truth table based approach

To full fill the requirement of replacement based approach first we required every gate replacement means it's clear that we need to replace every MVL gate equals to their binary gate. Here we can see with this an example; suppose we have one equality gate with us, it is in MVL form now we want to convert this gate in its equal binary gate. First we obtain MV truth table; we convert it into their BV truth table. Now we generate circuit using k-map, we got some circuit equation. Than we identify MSB and LSB input and output for each I/O. Now we can store it in library. We follow flow chart in Figure 4.2 to realize the process.

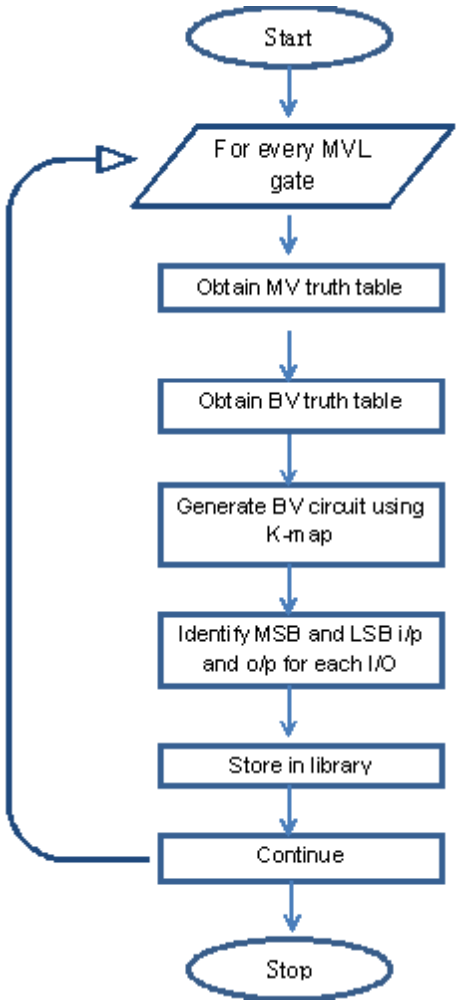

Fig. 4.2: Flow Chart of Gate Conversion

## 4.2 Design Entry of MVL Circuits

MVL circuits can be captured by our planned tool in schematic format. Tool should have following capabilities:

- ➢ It should be able to create input and output pins, with assignments of quaternary values.
- ➢ It should pick and place multivalued gates from the screen and assign values at inputs.
- ➢ Input should be able to connect require input and output of gate and change logic values accordingly.
- ➢ It should be able to save and restore the circuit.
- ➢ It should be able to edit the circuit.

➢ It should be able to display truth table of any gate.
➢ It should be able to convert and display binary circuit schematic.

Design entry and conversion facility should be planned in such a way so that one side should contain all gates and other side should contain input, output, editing features, conversion commands etc. one window should display multivalued design entry and other side should display the converted binary circuit.

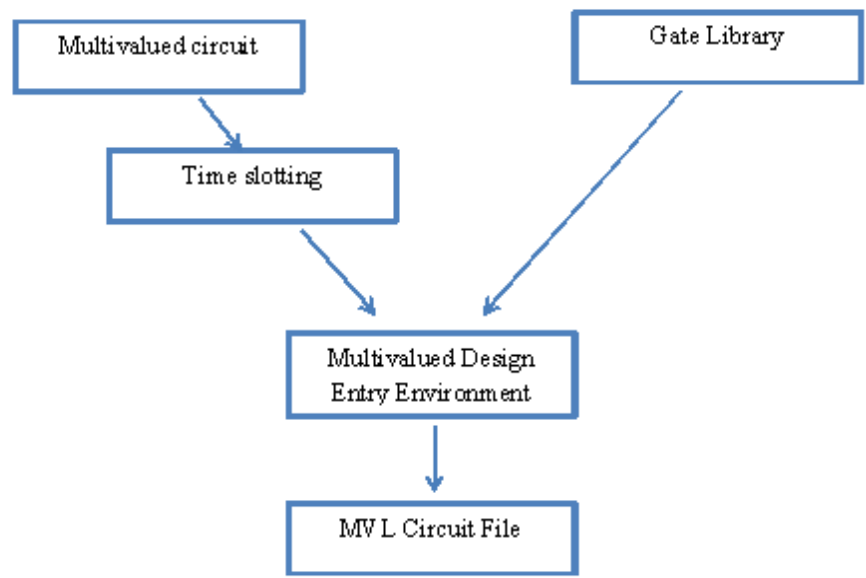

Fig. 4.3: MVL Circuit Design Entry

A designer has to prepare multivalued circuit in hand and divide the circuit operation/gates in timeslots. A timeslot is vertical portion of the circuit. There are two types of time slot namely gate slot and wiring slot. Within a gate slot at most one gate is available while horizontal viewing, whereas only wiring connectivity is shown in between gates within wiring slot. Figure 4.4(a) shows a MVL circuit and Figure 4.4 (b) shows corresponding slotted circuit.

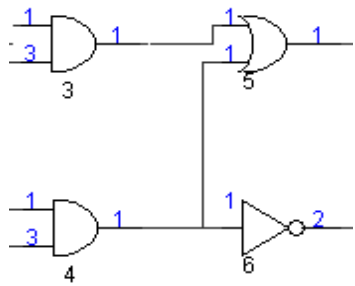 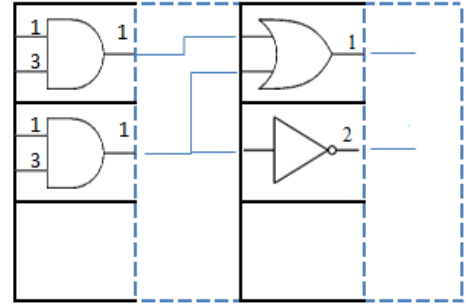

Fig. 4.4 (a): MVL circuit  Fig. 4.4 (b): MVL slotted circuit

Multivalued design entry environment takes slotted circuit and library in input to produce desired circuit interactively.

## 4.3  Algorithms

The tool operation involves design entry, conversion and saving of the circuit. Design entry process contains slot by slot circuit entry, editing if required. The circuit entry and editing the circuit are interactive processes. They involve schematic display and changing circuit accordingly. After desired circuit is entered and found correct, wiring connectivity algorithms connects input and output of the concerned gates. The circuit is converted into binary circuit and finally multivalued and binary circuit can be saved into a file. The above operations can be shown as flow chart Figure 4.5.
Each major step can further be divided into several smaller steps shown in 4.3.1.

### 4.3.1  Design Entry

Design entry involves circuit entry, editing and wiring connectivity. This being interactive process, object oriented analysis was done on entire system. The following major classes were identified for design entry purpose.

a) Gate class: - The purpose of this class is to enter, store and display any quaternary gate. This will contain structural information above connectivity and placement only.

Functionally related parameter will be available in derived classes only. The top compartment shows the class's name. The middle compartment lists the class's attributes. The bottom compartment lists the class's operations. When drawing a class element on a class diagram, you must use the top compartment, and the bottom two compartments are optional. Figure 4.6 shows Gate class.

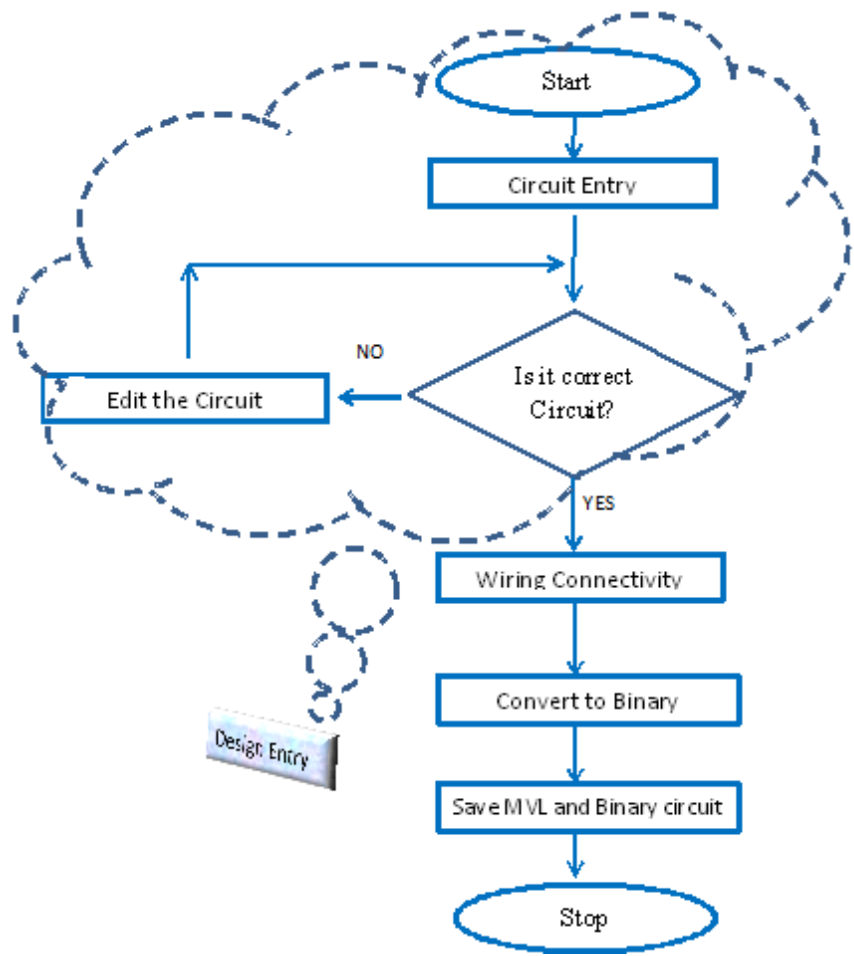

Fig. 4.5: Approach used for Schematic Entry

| Gate |
|---|
| I1:Input connection; |
| C: Coordinates |
| gateOutputCordinate: Coordinates |
| gateI1InputPosition: Coordinates |
| gateOutput : int |
| Timeline no: int |
| Slot no: int |
| Gate Type: int |
| Name: String |
| getname() |
| toString() |

Fig. 4.6: Class Diagram for Gate Class

Here we have few extended classes from Gate class like AND Class, NOT Gate shown in Fig 4.6 (a). In Gate class in take a input 1 only because this required according to gate input if gate required two input like AND gate than second input is derived in derived class which is extended from base class and if second input is not required like NOT class than it would be as like as but inherit the properties of base class.

| And Gate | | Not Gate |
|---|---|---|
| I2:Input connection; | | |
| id=0; | | id=0; |
| gateI2InputPosition: Coordinates | | |
| Gateprocessing() | | Gateprocessing() |
| PaintComponent() | | PaintComponent() |

Fig 4.6(a): Derived classes of Gate class

b) Binary Gate: - The purpose of this class is to store, display any binary gate corresponding to their respective quaternary gate. This will contain structural information above connectivity and placement only. Functionally related parameter will be available in derived class only. Figure 4.7 shows variable and methods of class. Some other classes (like Binary Conversion AND, Binary Conversion OR, Binary Conversion

MAX, MIN, Bit Comparator, Binary Decoder etc.) extends this class hierarchy. Here Binary input and binary output shows the hierarchy from base class.

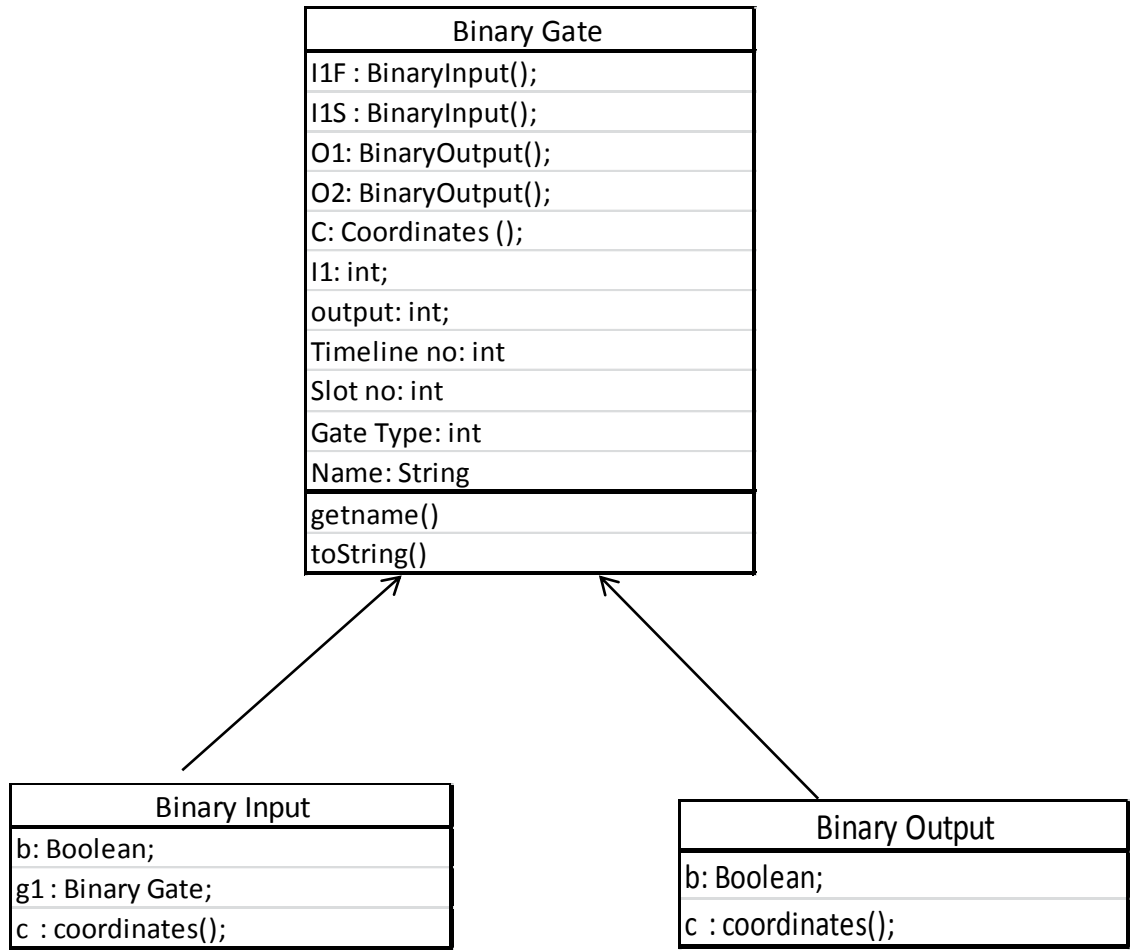

Fig. 4.7: Class diagram for Binary Gate Class

Here we have a few derived classes of Binary Gate class; we can see it in Figure 4.7(a). Like Binary AND gate class we have more classes to derived like OR, NOT, etc. which uses Binary gate class as a base class and extend it property in their own class. Here we can see that our input I2F and I2S is required in derived And class but in case Not Class it is not required.

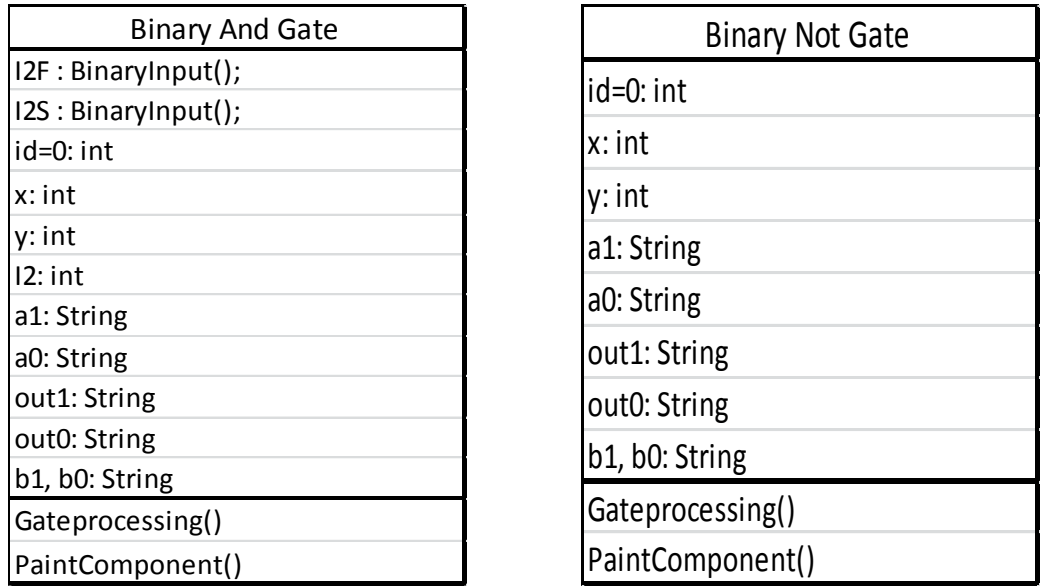

Fig 4.7 (a): Derived Class from Binary Gate Class

c) Coordinates: - The purpose of this class to store the coordinates on x axis and y axis. Fig. 4.8 shows variable of class. This class calls the coordinate and set parameters.

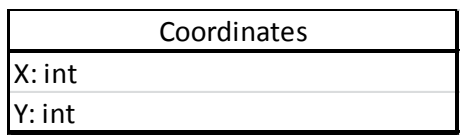

Fig. 4.8: Class Diagram for Coordinates Class

d) Gate List: - The purpose of this class to store gate, add gate, clear gate, remove and reconnect gate in vector data structure. This is also a part of super class. Fig. 4.9 shows variable and methods of class.

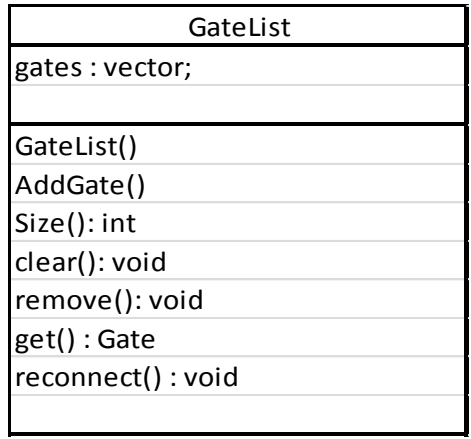

Fig. 4.9: Class diagram for Gate List Class

e) Wiring: - The purpose of this class to provide connectivity of wire between gates in diag. Fig. 4.10 shows variable and methods of class.

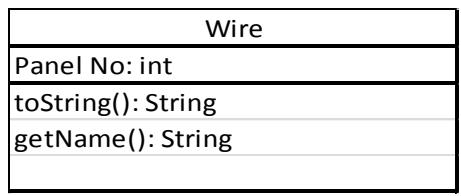

Fig. 4.10: class diagram for Wire Class

f) Circuit Line Coordinates: - The purpose of the circuit line coordinates class is to take input, output and difference of gates and to set their references between gates. Figure 4.11 shows variable and methods of class.

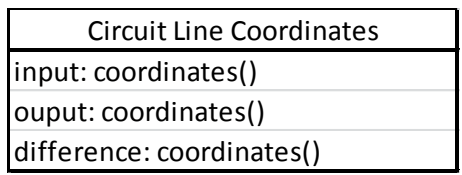

Fig. 4.11: Class Diagram for Circuit Line Coordinates

Once the classes have been identified, a GUI has been prepared to pick and place the gates and input objects. While placing the inputs property panels require input value for input type object and input connection for gate type object and this will be store as a parameter in appropriate class. There is algorithm which shows the design entry steps.

| | |
|---|---|
| **Algorithm 4.1:** Design Entry Algorithm | |
| INPUT: MVL Circuit | |
| OUTPUT: BVL Circuit | |

// MVL circuits are divided in time slots (TL) such that TL=0 input, TL=1 wiring, TL=2 gate, TL=3 wiring and last slot output.

// Schematic area is divided slot wise.

1. Initialize GUI and display
2. Initialize circuit array list
    // circuit array list contain gateItems (gate, inpin, outpin, and its parameter)
3. Read mouse click and event source
4. Case event source:
5. Select Input:      Display input at leftmost corner.
    5.1                Auto-generate number
    5.2                Assign input value from dialog box
    5.3                Store in circuit array list
    5.4                Break;
6. Select Output:     Display output at left most corner;
    6.1                Auto generate number
    6.2                Store in circuit array list
    6.3                Break;
7. Select Gate:       Capture gate and its input
    7.1                Display at left most corner
    7.2                Auto generate number
    7.3                Store in circuit array list
    7.4                Break;
8  Select moveToSlot: Select Displayed item move to appropriate slot and place
    8.1                Break;
9  Case Wiring:       Call Wiring algorithm
    9.1                Break;

| 10 | Case Convert: | Call Convert algorithm |
|---|---|---|
| 11 | Case Delete Gate: | Delete gate |
| | 11.1 | Break; |
| 12 | Case Delete Input: | Delete input |
| | 12.1 | Break; |
| 13 | Case Delete Output: | Delete output |
| | 13.1 | Break; |
| 14 | Case Open File: | Open and display circuit from file |
| | 14.1 | Break; |
| 15 | Save circuit: | MVL & BVL |
| | 15.1 | Break; |
| 16 | Case exit: | Exit |
| 17 | End switch | |
| 18 | Continue forever | |
| 19 | End | |

After the above algorithm design entry algorithm will perform. Now we will show further wiring connectivity algorithm and convert algorithm.

### 4.3.2 Wiring Connectivity

In order to diagrammatically show the connectivity a menu in panel has been planned to provide gate connectivity drawing, a gate list class and objects contains the list of gates entered in the design. The design is divided into timeslot and time line. Each time line and slot will contain display area and connectivity area duly defined through coordinates.

The connections are established as per the following algorithm. There will be a wiring/draw button in panel which activate this algorithm.

**Algorithm 4.2:** Wiring

INPUT: MVL Circuit with Gate

OUTPUT: Interconnected MVL Circuit

// Assuming MVL circuit is complete and store in time slotted manner.

// Timeslot TL=0 is input, TL=1 wiring, TL=2 gate, TL=3 wiring, TL=4 gate and last slot is output.

1. Capture MVL circuit and its data structure
2. For Time slot TL=2 to last TL increment by 2
3.    Extract gateItems list from current slot Circuit array list
4.      For every gateItem in current slot
5.        For every input connection of current slot gates
6.          Obtain input source
7.            Compute fan-out of input source
8.              Compute wiring path and display
9. Continue till last timeslot is over
10. End

After above algorithm connections are displayed.

### 4.3.3 Conversion Algorithm

This is our third phase where schematic design entry action performs that is conversion of circuit. Conversion algorithms convert multivalued circuit to binary valued circuit. This algorithm conversion based on replacement based algorithm. It takes input in multivalued gates and conversion it into their respective binary gates.

| **Algorithm 4.3:** Convert MVL to BVL |
|---|
| INPUT: MVL Circuit Gates and Connectivity |
| OUTPUT: BVL Circuit |

// Gate Item can be gate or input or output

// Assuming MVL circuit is complete and store in time slotted manner.

// Timeslot TL=0 is input, TL=1 wiring, TL=2 gate, TL=3 wiring, TL=4 gate and last slot is output.

// conversion library contains BVL gates with MSB and LSB input and output points.

1. Capture MVL circuit and its data structure.
2. Time slot TL=0 to last TL increment by 2
3. Extract gate item in current slot
4.     For every gate item in current slot.
    - 4.1     Obtain corresponding binary valued gate item
    - 4.2     Display and update BVL data structure with MSB & LSB input/output points.
5. Continue till last slot output display

    // for wiring connectivity in BVL circuit
6. Time slot TL=2 to last TL increment by 2
7. Extract gateItems list of current slot from circuit array list
8. For every gateItems in current slot
9. For every input connection of current slot gate item
    - 9.1   Obtain MSB and LSB input of gate item and its source item.
    - 9.2   Connect MSB input of gate item to MSB input of source item
    - 9.3   Connect LSB input of gate item to LSB input of source item
10. Continue till last gate item of last time slot.
11. End

The chapter worked out on tool development plan and algorithms to be implemented. User interface and result obtained will show & discussed in next chapter.

# Chapter 5

# IMPLEMENTATION AND RESULTS

After developing a tool name "MVL-DEV " the graphical representation and testing results are describe in this chapter.

## 5.1  GUI Interface for development and editing of Multivalued Circuit

This GUI interface will allow you to develop the multivalued circuit and edit them as well as allow us to set the properties of every gate and convert it in their respective binary conversion.

### 5.1.1  Hardware and Software Platform

For making the tool we use to hardware and software platform for that we use:

**Processor:** Intel(R) Core(TM) i3 CPU M 370 @ 2.40GHz, 2394 Mhz, 2 Core(s), 4 Logical Processor(s)

**RAM:** 4 GB

**OS platform:** Microsoft Windows 7 Ultimate, 64 bit OS

**Software Platform:** Net Beans IDE 7.0.1

**Language:** Java 1.6.0; Java Hot Spot(TM) Client VM 1.6.0-b105

### 5.1.2  Developing and Simulating Multivalued Circuit.

**A:** (Inputs) for developing and simulating the circuit first we require taking the input whether it can be (0, 1, 2, and 3) any of them and you take multiple inputs together.

**B:** (Circuit Drawing Sheet) when we select the inputs then it will draw on circuit drawing sheet which is middle of the tool. The input gate will draw in top left corner by default and then we will move it in any of the time slot in which we require. The partition of this window is done by beauty of art. It representation is very splendor. When gate is draw in top left corner than we move it, by mouse in any time slot or first time slot. Likewise we can draw multiple of gates here. The whole circuit will draw in this window which would be converted further in next window.

**C:** (Logic Gates) this section known as Logic Gates, here total ten gates are showing, name are given like this AND, OR, NOT, BITSWAP, XOR, INWARD INVERTER, OUTWARD INVRTER, EQULITY, MIN, MAX. When we click on any of gates it will display in top left corner of drawing sheet by default, now we move these gate in time lines.so we can add multiple gates together.

**D:** (Properties) this section shows the name properties. This will help us in to set the property of every input and gates. It gives the idea that how we can set every input value of every gate and save their values in data structure. So this section have some other section E, F, X, W, U, V, T which we will discuss in there turn.

**E:** (Input A) Input A require input reference of section A or input gates.

**F:** (Input B) Input B requires input reference of section A or input gates.

**G:** (Table) this section will show the truth table of every gate.

**H:** (Output) the output will give the final output according to truth-table combination and will display it into the output.

**I:** (Delete) by selecting gate Delete button will delete the gate from data structure and help us to add other gate on their respective position.

**J:** (Draw) when we click on draw button than on drawing sheet wires will be connected between gates ad input and outputs according to their input references.

**K:** (Convert) Quaternary circuit we will draw on circuit drawing sheet that will convert into binary circuit with replacement of every gate.

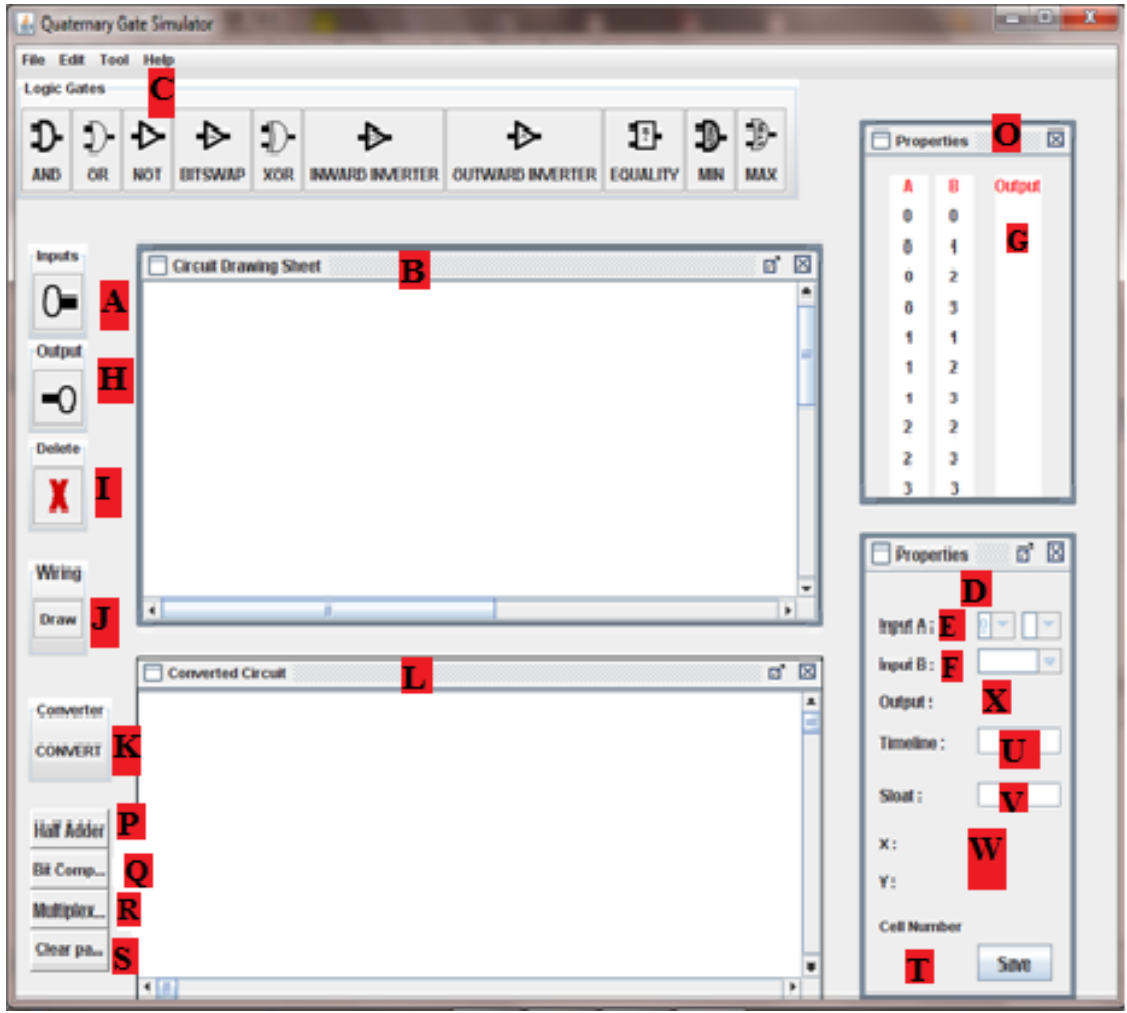

Fig. 5.1: "MVL dev. tool representation"

**L:** (Converted circuit) this is second display window where we will see the converted binary circuit of quaternary circuit. This is our output circuit and we get conclusion here.

**O:** (Properties) this panel draws the truth table of every selected gate in given time lining of circuit drawing sheet.

**P:** (Half adder) there are some standard circuit which I added in this tool which will show there truth table (Quaternary, Binary both) and binary conversion of Half adder.

**Q:** (Bit comparator) it will show the bit comparator circuit and there truth table (Quaternary, Binary both).

**R:** (Multiplexer) it will show 4*1 MUX conversion binary circuit and their truth table (Quaternary, Binary both).

**S:** (Clear panel) this button will clear the converted circuit sheet.

**T:** (Save) save button save the input values and properties in data structure.

**U:** (Time Line) gate location in time line.

**V:** (Slot) slot location in given time line, and gate location too.

**X, Y:** (Axis) Axis location in drawing sheet window.

### 5.1.3 User Session

In this session we will show the steps followed by the user of tool to enter the circuit, edit, convert & save the circuit. User has to following steps:

1). Circuit preparation: - The user has kept ready to circuit schematic. Schematic has divided into timeslot. In this user session we assume following circuit is shown in Figure 5.2 is taken by user and it's a time slotted version is also prepared as shown in Figure 5.2.

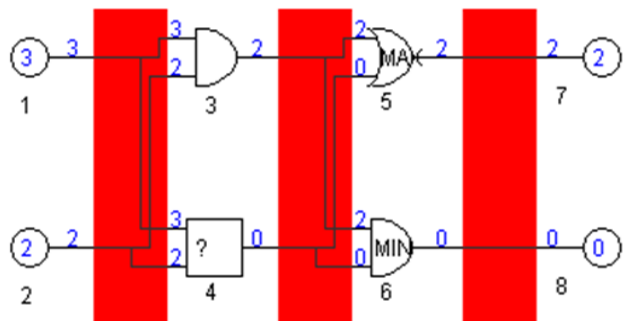

Fig. 5.2: MVL Circuit

2). Entering the circuit: - For entering the circuit first we take the input values after selecting input button, it could be any between 0 to 3 than input will display in top left corner of the display window. Every selected input, output and gates first will display in this top left corner block. Now we will drag through mouse that input in timeline and slots. Now we can add more gates in this circuitry. By selecting each single gates/ inputs we can set their properties, also set their input references of every gate. After adding each gate in circuitry and set their properties we can click on draw button that will draw the wiring connectivity of circuit according to input references. A circuit diagram is shown in Figure 5.2.

3). Editing in circuit: - Suppose we want to change a gate no 3. We want to change this gate by replacing another gate than we will simply click on gate no 3 (AND Gate), and we have delete button in tool to delete the gate by simple click. After that we will add another gate against this. And change the property of gate and save to it. We can see here in below diagram edited circuit.

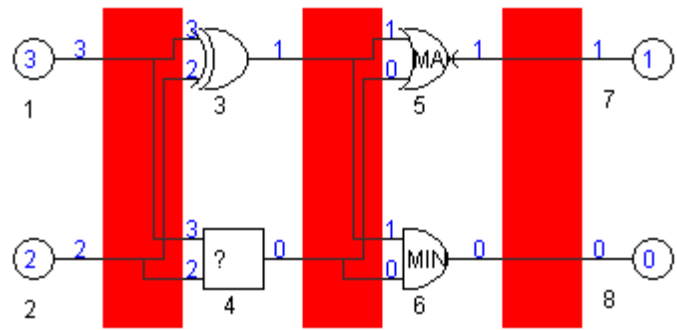

Fig. 5.3: Edited Multivalued Circuit of Fig.5.2

4). Drawing / wiring circuit: - In tool a draw button is placed, which is used to draw wiring connectivity between gates. It will catch the references of gate and draw the connection of wire as shown in above two Figure 5.2 and 5.3.

5). Conversion: - In tool a convert button is placed which will convert the multivalued circuit to binary circuit. In this conversion we have every multivalued gate conversion to binary conversion. Every gate has their replacement gates in binary. A converted circuit we can see in below Figure 5.4. Conversion algorithm shows in chapter 4 at algorithm 4.3.

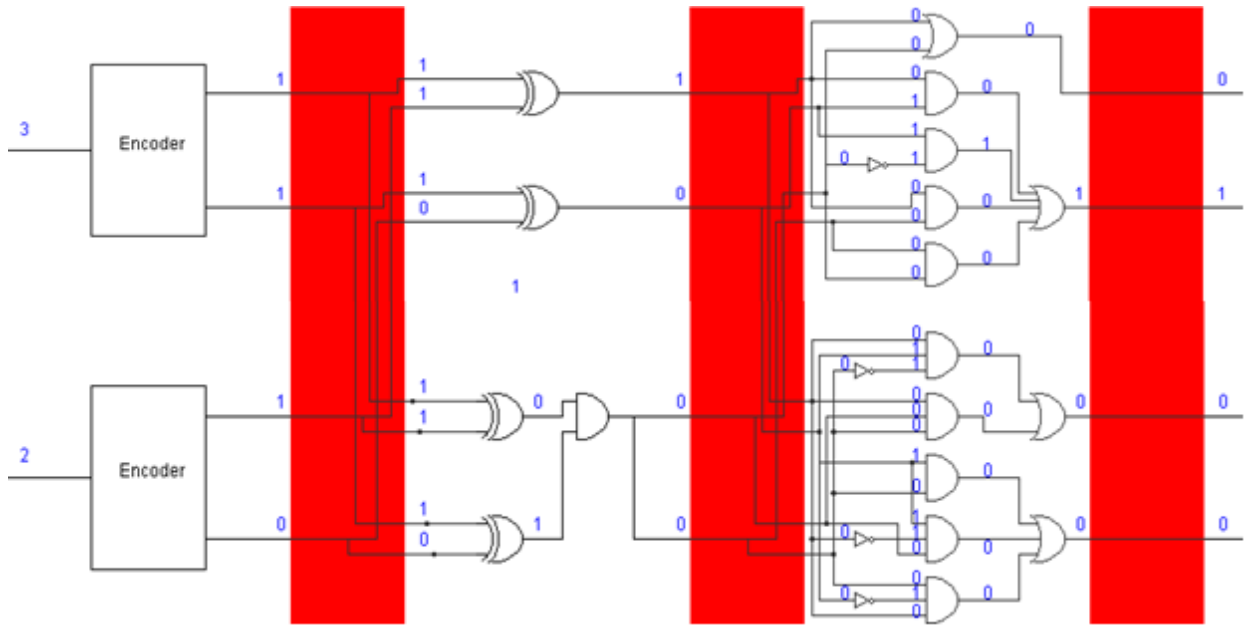

Fig. 5.4: Converted Circuit of Multivalued Circuit to Binary

## 5.2 Testing and Verification

This section included results of some testing for purpose of verification.

### 5.2.1 Gate Level Testing

Gate level testing incudes testing of simulation results of single gate, this can be done by adding only single gate to the circuit and simulate that circuit with single gate and verifying the outcome of circuit. If the produced output (truth table of added gate) matches with actual truth table of gate the test result is positive (test is successful) else negative. The Table 5.1 gives the verification result for all individual gates used in this dissertation.

Table 5.1: Results of Gate Level Testing

| Sr No | Test case | Verification result | Test result |
|---|---|---|---|
| 1 | AND Gate | Result verified | Test Successful |
| 2 | OR Gate | Result verified | Test Successful |
| 3 | NOT Gate | Result verified | Test Successful |
| 4 | XOR Gate | Result verified | Test Successful |
| 5 | BIT SWAP Gate | Result verified | Test Successful |
| 6 | INWARD INVERTER Gate | Result verified | Test Successful |
| 7 | OUTWARD INVERTER Gate | Result verified | Test Successful |
| 8 | EQUILITY Gate | Result verified | Test Successful |
| 9 | MAX Gate | Result verified | Test Successful |
| 10 | MIN Gate | Result verified | Test Successful |

### 5.2.2 Circuit Level Testing

Circuit level testing includes testing of a circuit with their equivalent conversion in binary. We already discussed few multivalued circuit in chapter 3.here we have their converted circuit in binary and their truth table so that we can prove their circuit level testing.

**1).** Quaternary 1 to 4 decoder: Here we have one multivalued circuit and its truth table. This

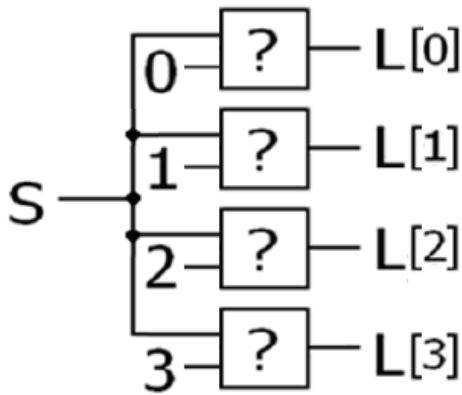

| S | Decoder (L) | | | |
|---|---|---|---|---|
| | [0] | [1] | [2] | [3] |
| 0 | 3 | 0 | 0 | 0 |
| 1 | 0 | 3 | 0 | 0 |
| 2 | 0 | 0 | 3 | 0 |
| 3 | 0 | 0 | 0 | 3 |

Fig. 5.5: Quaternary 1 to 4 decoder     Table 5.2: Truth table of decoder

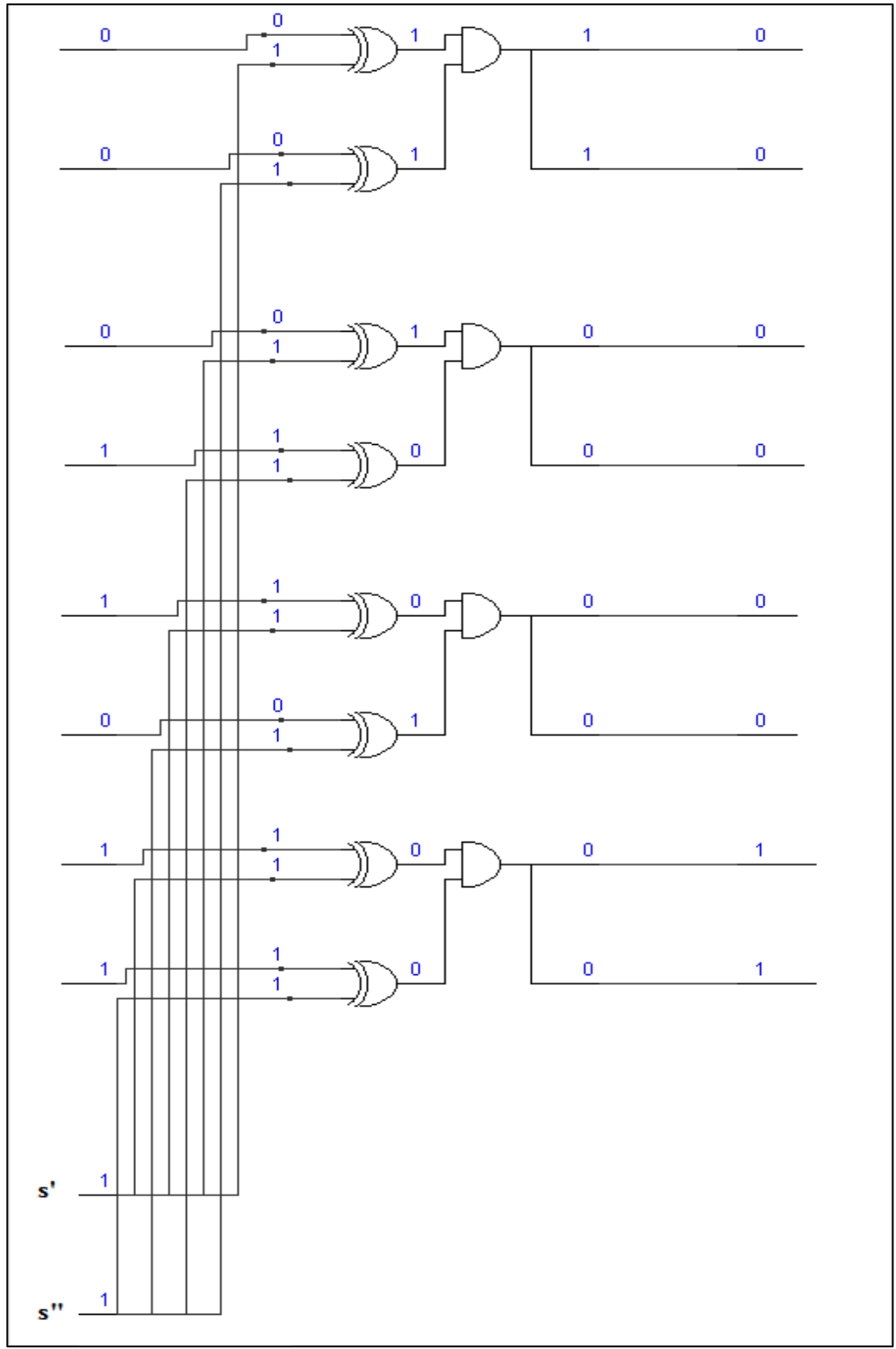

Fig. 5.6: Converted Binary Circuit of Quaternary 1 to 4 decoder

Table 5.3: Binary Truth Table of Quaternary 1 to 4 Decoder

| S'S" | Decoder (L) | | | |
|---|---|---|---|---|
| | $L_0[0], L_1[0]$ | $L_0[0], L_1[1]$ | $L_0[1], L_1[0]$ | $L_0[1], L_1[1]$ |
| 0,0 | 1,1 | 0,0 | 0,0 | 0,0 |
| 0,1 | 0,0 | 1,1 | 0,0 | 0,0 |
| 1,0 | 0,0 | 0,0 | 1,1 | 0,0 |
| 1,1 | 0,0 | 0,0 | 0,0 | 1,1 |

A quaternary circuit 1-to-4 decoder has one input and four outputs. Only one of the outputs can be equal to 3 at a time, all other outputs remain 0 as Figure 5.5 shows a Quaternary circuit and its equivalent Table 5.2 in quaternary. In Figure 5.6 binary equivalent circuit is shown, here we give input 3 on selection lines S' and S" and get an output on 3. Tool will convert the multivalued circuit into binary circuit. Table 5.3 shows there binary equivalent of Table 5.2.

2). Arbitrary Circuits:

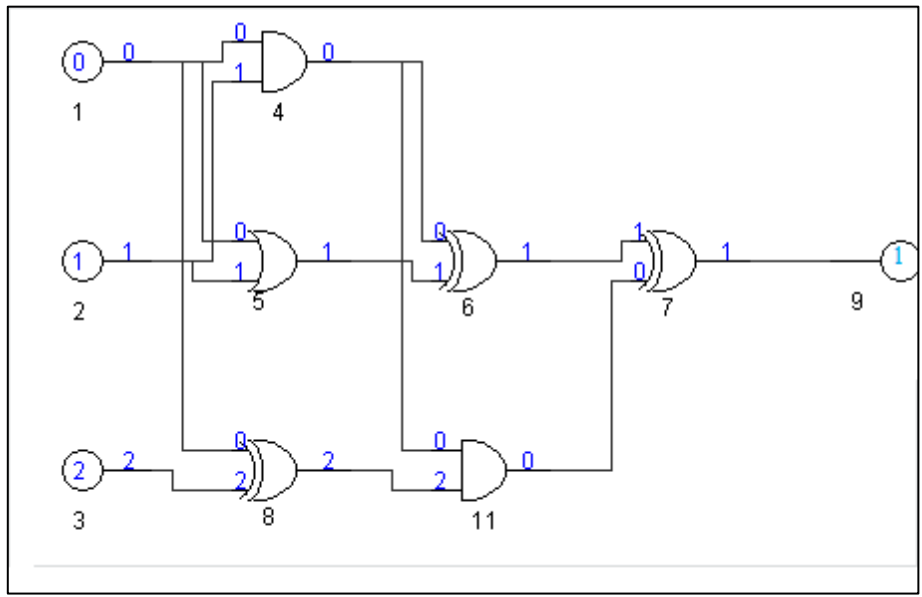

Fig. 5.7: Tool Generated Circuit

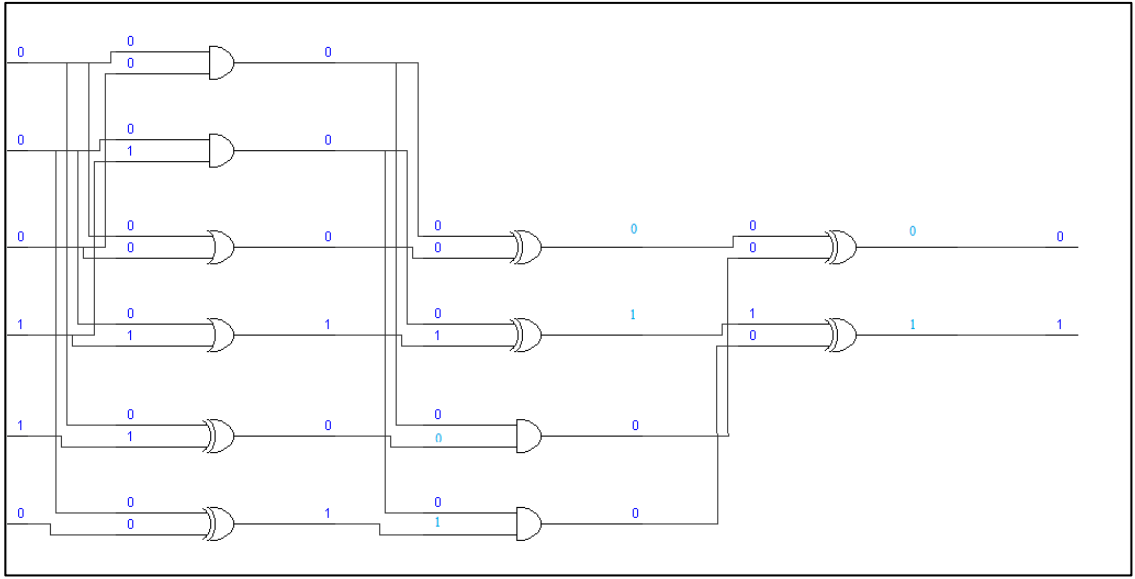

Fig. 5.8: Output of Tool Generated Circuit in Binary Valued

Table 5.4: MVL Truth Table of Arbitrary Circuit (Figure5.7)

| A | B | C | y |
|---|---|---|---|
| 0 | 0 | 0 | 0 |
| 0 | 0 | 1 | 0 |
| 0 | 0 | 2 | 0 |
| 0 | 0 | 3 | 0 |
| 0 | 1 | 0 | 1 |
| 0 | 1 | 1 | 1 |
| 0 | 1 | 2 | 1 |
| 0 | 1 | 3 | 1 |
| 0 | 2 | 0 | 2 |
| 0 | 2 | 1 | 2 |
| 0 | 2 | 2 | 2 |
| 0 | 2 | 3 | 2 |
| 0 | 3 | 0 | 3 |
| 0 | 3 | 1 | 3 |
| 0 | 3 | 2 | 3 |
| 0 | 3 | 3 | 3 |

| A | B | C | y |
|---|---|---|---|
| 1 | 0 | 0 | 1 |
| 1 | 0 | 1 | 1 |
| 1 | 0 | 2 | 1 |
| 1 | 0 | 3 | 1 |
| 1 | 1 | 0 | 0 |
| 1 | 1 | 1 | 1 |
| 1 | 1 | 2 | 0 |
| 1 | 1 | 3 | 0 |
| 1 | 2 | 0 | 3 |
| 1 | 2 | 1 | 3 |
| 1 | 2 | 2 | 3 |
| 1 | 2 | 3 | 3 |
| 1 | 3 | 0 | 2 |
| 1 | 3 | 1 | 3 |
| 1 | 3 | 2 | 2 |
| 1 | 3 | 3 | 2 |

| A | B | C | y |
|---|---|---|---|
| 2 | 0 | 0 | 2 |
| 2 | 0 | 1 | 2 |
| 2 | 0 | 2 | 2 |
| 2 | 0 | 3 | 2 |
| 2 | 1 | 0 | 3 |
| 2 | 1 | 1 | 3 |
| 2 | 1 | 2 | 3 |
| 2 | 1 | 3 | 3 |
| 2 | 2 | 0 | 0 |
| 2 | 2 | 1 | 2 |
| 2 | 2 | 2 | 0 |
| 2 | 2 | 3 | 0 |
| 2 | 3 | 0 | 3 |
| 2 | 3 | 1 | 3 |
| 2 | 3 | 2 | 1 |
| 2 | 3 | 3 | 1 |

| A | B | C | y |
|---|---|---|---|
| 3 | 0 | 0 | 3 |
| 3 | 0 | 1 | 3 |
| 3 | 0 | 2 | 3 |
| 3 | 0 | 3 | 3 |
| 3 | 1 | 0 | 3 |
| 3 | 1 | 1 | 2 |
| 3 | 1 | 2 | 3 |
| 3 | 1 | 3 | 2 |
| 3 | 2 | 0 | 3 |
| 3 | 2 | 1 | 3 |
| 3 | 2 | 2 | 1 |
| 3 | 2 | 3 | 1 |
| 3 | 3 | 0 | 3 |
| 3 | 3 | 1 | 2 |
| 3 | 3 | 2 | 1 |
| 3 | 3 | 3 | 0 |

Here we take typical arbitrary circuit which has three quaternary variable inputs. Its multivalued truth table is shown in Table 5.4 in four parts. Its equivalent binary converted circuit is in Figure 5.8. Truth table of binary equivalent conversion is shown in Table 5.5. we can see here by circuit and his truth table that there is so much complexity in binary circuit and his truth table when we have multiple input. Here we have 3 variables and 4 inputs; possible number of combination made in this is 64. We already shown in Table 5.4

Table 5.5: Truth Table of Equivalent Binary Conversion

| a1 | a0 | b1 | b0 | c1 | c0 | y1 | y0 |
|---|---|---|---|---|---|---|---|
| 0 | 0 | 0 | 0 | 0 | 0 | 0 | 0 |
| 0 | 0 | 0 | 0 | 0 | 1 | 0 | 0 |
| 0 | 0 | 0 | 0 | 1 | 0 | 0 | 0 |
| 0 | 0 | 0 | 0 | 1 | 1 | 0 | 0 |
| 0 | 0 | 0 | 1 | 0 | 0 | 0 | 1 |
| 0 | 0 | 0 | 1 | 0 | 1 | 0 | 1 |
| 0 | 0 | 0 | 1 | 1 | 0 | 0 | 1 |
| 0 | 0 | 0 | 1 | 1 | 1 | 0 | 1 |
| 0 | 0 | 1 | 0 | 0 | 0 | 1 | 0 |
| 0 | 0 | 1 | 0 | 0 | 1 | 1 | 0 |
| 0 | 0 | 1 | 0 | 1 | 0 | 1 | 0 |
| 0 | 0 | 1 | 0 | 1 | 1 | 1 | 0 |
| 0 | 0 | 1 | 1 | 0 | 0 | 1 | 1 |
| 0 | 0 | 1 | 1 | 0 | 1 | 1 | 1 |
| 0 | 0 | 1 | 1 | 1 | 0 | 1 | 1 |
| 0 | 0 | 1 | 1 | 1 | 1 | 1 | 1 |

| a1 | a0 | b1 | b0 | c1 | c0 | y1 | y0 |
|---|---|---|---|---|---|---|---|
| 0 | 1 | 0 | 0 | 0 | 0 | 0 | 1 |
| 0 | 1 | 0 | 0 | 0 | 1 | 0 | 1 |
| 0 | 1 | 0 | 0 | 1 | 0 | 0 | 1 |
| 0 | 1 | 0 | 0 | 1 | 1 | 0 | 1 |
| 0 | 1 | 0 | 1 | 0 | 0 | 0 | 0 |
| 0 | 1 | 0 | 1 | 0 | 1 | 0 | 1 |
| 0 | 1 | 0 | 1 | 1 | 0 | 0 | 0 |
| 0 | 1 | 0 | 1 | 1 | 1 | 0 | 0 |
| 0 | 1 | 1 | 0 | 0 | 0 | 1 | 1 |
| 0 | 1 | 1 | 0 | 0 | 1 | 1 | 1 |
| 0 | 1 | 1 | 0 | 1 | 0 | 1 | 1 |
| 0 | 1 | 1 | 0 | 1 | 1 | 1 | 1 |
| 0 | 1 | 1 | 1 | 0 | 0 | 1 | 0 |
| 0 | 1 | 1 | 1 | 0 | 1 | 1 | 1 |
| 0 | 1 | 1 | 1 | 1 | 0 | 1 | 0 |
| 0 | 1 | 1 | 1 | 1 | 1 | 1 | 0 |

| a1 | a0 | b1 | b0 | c1 | c0 | y1 | y0 |
|---|---|---|---|---|---|---|---|
| 1 | 0 | 0 | 0 | 0 | 0 | 1 | 0 |
| 1 | 0 | 0 | 0 | 0 | 1 | 1 | 0 |
| 1 | 0 | 0 | 0 | 1 | 0 | 1 | 0 |
| 1 | 0 | 0 | 0 | 1 | 1 | 1 | 0 |
| 1 | 0 | 0 | 1 | 0 | 0 | 1 | 1 |
| 1 | 0 | 0 | 1 | 0 | 1 | 1 | 1 |
| 1 | 0 | 0 | 1 | 1 | 0 | 1 | 1 |
| 1 | 0 | 0 | 1 | 1 | 1 | 1 | 1 |
| 1 | 0 | 1 | 0 | 0 | 0 | 0 | 0 |
| 1 | 0 | 1 | 0 | 0 | 1 | 1 | 0 |
| 1 | 0 | 1 | 0 | 1 | 0 | 0 | 0 |
| 1 | 0 | 1 | 0 | 1 | 1 | 0 | 0 |
| 1 | 0 | 1 | 1 | 0 | 0 | 1 | 1 |
| 1 | 0 | 1 | 1 | 0 | 1 | 1 | 1 |
| 1 | 0 | 1 | 1 | 1 | 0 | 0 | 1 |
| 1 | 0 | 1 | 1 | 1 | 1 | 0 | 1 |

| a1 | a0 | b1 | b0 | c1 | c0 | y1 | y0 |
|---|---|---|---|---|---|---|---|
| 1 | 1 | 0 | 0 | 0 | 0 | 1 | 1 |
| 1 | 1 | 0 | 0 | 0 | 1 | 1 | 1 |
| 1 | 1 | 0 | 0 | 1 | 0 | 1 | 1 |
| 1 | 1 | 0 | 0 | 1 | 1 | 1 | 1 |
| 1 | 1 | 0 | 1 | 0 | 0 | 1 | 1 |
| 1 | 1 | 0 | 1 | 0 | 1 | 1 | 0 |
| 1 | 1 | 0 | 1 | 1 | 0 | 1 | 1 |
| 1 | 1 | 0 | 1 | 1 | 1 | 1 | 0 |
| 1 | 1 | 1 | 0 | 0 | 0 | 1 | 1 |
| 1 | 1 | 1 | 0 | 0 | 1 | 1 | 1 |
| 1 | 1 | 1 | 0 | 1 | 0 | 0 | 1 |
| 1 | 1 | 1 | 0 | 1 | 1 | 0 | 1 |
| 1 | 1 | 1 | 1 | 0 | 0 | 1 | 1 |
| 1 | 1 | 1 | 1 | 0 | 1 | 1 | 0 |
| 1 | 1 | 1 | 1 | 1 | 0 | 0 | 1 |
| 1 | 1 | 1 | 1 | 1 | 1 | 0 | 0 |

Hence we observe our tool is able to perform design entry editing and conversion and saving task correctly. It will be useful for multivalued design entry and interfacing with binary logic based system.

**Chapter 6**

# CONCLUSION AND FUTURE SCOPE

In this dissertation we plan a multivalued circuit processing environment that it offers ease of multivalued circuit design and development platform of binary valued system.

Looking to complexity of today's circuit (i.e. multi millennium gate size circuit), it is important that to develop circuits at higher than binary logic system so that design complexity is reduced. However implementation platform available are only in binary logic system.

We have developed a tool" MVL-DEV" that can be useful for developers for generating and testing multivalued circuits. Our specific contribution in this dissertation by developing "Multivalued circuit processing environment" that can be

- ✓ Creation of standard libraries in MVL.
- ✓ Simulate the developed multivalued circuit to view output of quaternary circuit and truth table of gates.
- ✓ Editing in multivalued circuit.
- ✓ Drawing wiring connectivity in multivalued circuit.
- ✓ Conversion of quaternary to binary circuit.
- ✓ Saving the generated circuit.
- ✓ Providing standard circuits as ready example design capture.

**Future Scope**

The developed tool is now able to capture multivalued design entry and provide binary valued logic design. This tool can be further improved by:

- ✓ Saving in more than one format.
- ✓ This can also extended for converting the design in reversible/quantum circuit.
- ✓ Further extension is possible by interfacing with hardware description languages.
- ✓ It may also be extended to take ternary and other multivalued design formats.

# BIBLIOGRAPHY

[1] Hurst, S.L, "Multiple-Valued Logic; its Status and its Future", Computers, IEEE Transactions on, **33** (12), Dec. 1984, pp.1160-1179.

[2] Miller, D.M. Thornton, and M.A. "Multiple Valued Logic: Concepts and Representations", Synthesis Lectures on Digital Circuits and Systems, Morgan & Claypool Publishers, 2007.

[3] Smith, K.C., "The Prospects for Multivalued Logic- A Technology and Applications View", Computers, IEEE Transactions on, 30, pp. 619-634, 1981.

[4] Fijany, A., Vatan, F.Mojarradi, M., Toomarian, B. Blalock, B.Akarvardar, K. Cristoloveanu, S. and Gentil, P. "The G4-FET- a universal and programmable logic gate", Proceedings of the 15th ACM Great Lakes symposium on VLSI (GLSVLSI2005), pp. 349-352, April 2005.

[5] Keshavarzian, A.P. and Navi, K., "Optimum Quaternary Galois Field Circuit Design through Carbon Nano Tube Technology", Proceedings, International Conference on Advanced Computing and Communications (ADCOM 2007), pp. 214-219, Dec 2007.

[6] Chattopadhyay, T. Taraphdar, C. Roy, and J.N, "Quaternary Galois field adder based all-optical multivalued logic circuits", Applied Optics, 48 (22), pp. E35-E44, 2009.

[7] Khan, and M.H.A., "Reversible Realization of Quaternary Decoder, Multiplexer, and Demultiplexer Circuits", Engineering Letters, 15 (2), pp. 203-207, 2007.

[8] Falkowski, B.J., and Rahardja S. "Generalized hybrid arithmetic canonical expansions for completely specified quaternary functions", Electronics Letters, 37 (16), pp. 1006-1007, Aug. 2001.

[9] Khan, M.H.A., Siddika, N.K. Perkowski, and M.A.; "Minimization of Quaternary Galois Field Sum of Products Expression for Multi-Output Quaternary Logic Function Using Quaternary Galois Field Decision Diagram", Proceedings, 38th IEEE International Symposium on Multiple-Valued Logic (ISMVL 2008), pp. 125-130, May 2008.

[10] Khan, M.M.M., Biswas, A.K., Chowdhury, S. Tanzid, M., Mohsin, K.M., Hasan, M., Khan, and A.I. "Quantum realization of some quaternary circuits", Proceedings, TENCON 2008, IEEE Region 10 Conference, Nov. 2008.

[11] Yasuda, Y., Tokuda, Y., Zaima, S., Pak, K., Nakamura, T. and Yoshida A., "Realization of quaternary logic circuits by n-channel MOS devices", Solid-State Circuits, IEEE Journal of, 21 (1), pp. 162-168, Feb. 1986.

[12] Park S.J., Yoon B.H., Yoon K.S., Kim H.S., "Design of quaternary logic gate using double pass-transistor logic with neuron MOS down literal circuit", Proceedings, 34th IEEE International Symposium on Multiple-Valued Logic (ISMVL 2004), pp.198-203, May 2004.

[13] Da Silva, R.C.G., Boudinov, H. and Carro L., "A novel Voltage-mode CMOS quaternary logic design", Electron Devices, IEEE Transactions on, 53 (6), pp. 1480, June 2006.

[14] Inaba, M., Tanno, K., Ishizuka, O. "Realization of NMAX and NMIN functions with multi-valued voltage comparators", Proceedings, 31st IEEE International Symposium on Multiple-Valued Logic (ISMVL 2001), pp. 27-32, May 2001.

[15] Datla, S.R., Thornton, and M.A., "Quaternary Voltage-Mode Logic Cells and Fixed-Point Multiplication Circuits", Proceedings, 40th IEEE International Symposium on Multiple-Valued Logic (ISMVL 2010), pp. 128-133, June 2010.

[16] Chattopadhyay, T. Roy and J.N, "Polarization-encoded all-optical quaternary multiplexer and demultiplexer - A proposal", Optik - International Journal for Light and Electron Optics, 120 (17), pp. 941-946, Nov. 2009.

[17] Bartelt, Terry. Wisconsin Technical College System."Wisc-Online".Basic Logic Gates. [http://www.wisc-online.com/Objects/ViewObject.aspx?ID=dig1302].

[18] Imperial College, London. Department of Computing. "Lecture 1: An Introduction to Boolean Algebra", [http://www.doc.ic.ac.uk/~dfg/hardware/HardwareLecture01.pdf].

[19] Huntington, and E.V., "Sets of Independent Postulates for the Algebra of Logic", Transactions of the American Mathematical Society, 5(3), pp. 288-309, July1904.

[20] X.W.Wu, "CMOS ternary logic circuits", IEEE PROCEEDINGS, Vol. 137, Pt. G, No. I, February 1990.

[21] Xiaoqiang Shou, Nader Kalantari, Michael M. Green, "Design of CMOS Ternary Latches", IEEE TRANSACTIONS ON CIRCUITS AND SYSTEMS—I: REGULAR PAPERS, VOL. 53, NO. 12, December 2006.

[22] Ion Profeanu, "A Ternary Arithmetic and Logic", ISBN: 978-988-17012-9-9, ISSN: 2078-0958 (Print); ISSN: 2078-0966 (Online); Proceedings of the World Congress on Engineering 2010 Vol. I , WCE 2010, June 30 - July 2, 2010, London, U.K.

[23] Kanchan S. Gorde, "Design and simulation of ternary logic based arithmetic circuits" issue 2, vol-1, april-june 2010.

[24] Afolayan A. Obiniyi, Ezugwu E. Absalom, Kwanashie Adako, "Arithmetic logic design with color coded ternary for ternary computing" International Journal of Computer Applications (0975 – 8887) Volume 26– No.11, July 2011.

[25] V.T.Gaikwad, and Dr. P. R. Deshmukh, "Multi-Valued Logic Applications in the Design of Switching Circuits" International Journal of Advanced Research in Computer Science and Software Engineering, Volume 2, Issue 5, ISSN: 2277 128X, May 2012.

[26] Noboru Takagi, and Kyoichi Nakashima, "Discrete Interval Truth Values Logic and its Application". IEEE TRANSACTIONS ON COMPUTERS, VOL. 49, NO.3, MARCH 2000.

[27] Craig M. Files and Marek A. Perkowski, "New Multivalued Functional Decomposition Algorithms Based on MDDs". IEEE TRANSACTIONS ON COMPUTER AIDED DESIGN OF INTEGRATED CIRCUITS AND SYSTEMS, VOL. 19, NO. 9, SEPTEMBER 2000.

[28] Sunil P. Khatri, Subarnarekha Sinha, Robert K. Brayton, and Alberto L. Sangiovanni-Vincentelli, "SPFD-Based Wire Removal in Standard-Cell and Network of PLA Circuits".IEEE TRANSACTIONS ON COMPUTER-AIDED DESIGN OF INTEGRATED CIRCUITS AND SYSTEMS, VOL. 23, NO. 7, JULY 2004.

[29] DušankaBundalo, ZlatkoBundalo, and BranimirDordevic, "Design of Quaternary Logic Systems and Circuits". ; FACTA UNIVERSITATIS (NIŠ) SER.: ELEC. ENERG. vol. 18, No. 1, April 2005, 45-56.

[30] Vasundara Patel K S, K S Gurumurthy, "Design of High Performance Quaternary Adders".International Journal of Computer Theory and Engineering, Vol.2, No.6, December, 2010, 1793-8201.

[31] Cristiano Lazzari, Paulo Flores, José Monteiro, and Luigi Carro "A New Quaternary FPGA Based on a Voltage-mode Multi-valued Circuit" 978-3-9810801-6-2/DATE10 © 2010.

[32] Vasundara Patel K S, and K S Gurumurthy, "Quaternary Sequential Circuits" IJCSNS International Journal of Computer Science and Network Security, VOL.10 No.7, July 2010.

[33] Vasundara Patel k s, and k s gurumurthy; "Arithmetic operation in multivalued logic "International journal of vlsi design & communication system (VLSCS), vol 1, no 1, march 2010.

[34] Sheng Lin, Yong-Bin Kim, and Fabrizio Lombardi, "Design of a Ternary Memory Cell Using CNTFETs" IEEE TRANSACTIONS ON NANOTECHNOLOGY, VOL. 11, NO. 5, SEPTEMBER 2012.

[35] Umesh Kumar and Rajiv Kapoor, "CMOS Body Driven Quaternary Logic Generator". IOSR Journal of VLSI and Signal Processing (IOSR-JVSP) ISSN: 2319 – 4200, ISBN No. : 2319 – 4197 Volume 1, Issue 1 (Sep-Oct. 2012), PP 46-50.

[36] Ifat Jahangir, Anindya Das and MasudHasan., "Formulation and Development of a Novel Quaternary Algebra"

[37] D. Etiemble and K. Navi, "Algorithms and multi-valued circuits for the multioperand addition in the binary stored-carry number system" 1993 IEEE 1063-6889/93.

[38] Elena Dubrova, "Multiple-Valued Logic in VLSI: Challenges and Opportunities"

[39] Robert K Brayton and Sunil P Khatri, "Multi-valued Logic Synthesis"

[40] Ashok Muthukrishnan and C. R. Stroud, "Multi-Valued Logic Gates for Quantum Computation" arXiv:quant-ph/0002033v2 12 Jun 2000.

[41] A. Morgül and TurgayTemel, "Level Restoration for Multivalued Logic Circuit".

[42] Hemant Dhabhai, Abhishek Katariya, Geetam Tomar, Yogesh Krishna, "Optimization of Binary and Multivalued Digital Circuit using MVSIS and AIG rewriting" Journal of Electronic and Electrical Engineering ,ISSN: 0976–8106 & E-ISSN: 0976–8114, Vol. 2, Issue 1, pp-24-29, 2011.